\documentclass[%
 reprint,
 amsmath,amssymb,
 aps,
]{revtex4-2}

\usepackage{graphicx}
\usepackage{dcolumn}
\usepackage{bm}
\usepackage{siunitx}
\usepackage{mhchem}
\usepackage{color,soul}
\usepackage[colorlinks=true,citecolor=blue,urlcolor=blue]{hyperref}
\usepackage{enumitem}
\usepackage{mhchem}
\usepackage{mathtools}

\usepackage{newtxtext,newtxmath}
\DeclareMathAlphabet\mathbfcal{OMS}{cmsy}{b}{n}

\newcommand{\BiSe}{\ce{Bi2Se3} }
\newcommand{\BiSenospace}{\ce{Bi2Se3}}
\newcommand{\BField}{\mathcal{B}}
\newcommand{\Bparam}{B}
\newcommand{\magneticLength}{l_{\BField}}
\newcommand{\FermiVelocity}{v_{\mathrm{F}}}
\newcommand{\iu}{\mathrm{i}}
\newcommand{\ec}{\mathrm{e}}
\renewcommand{\vec}[1]{\boldsymbol{#1}}

\begin{document}

\preprint{APS/123-QED}

\title{Edge states of a \texorpdfstring{\BiSe}{Bi2Se3 }nanosheet in a perpendicular magnetic field}

\author{S.P.J. Koenis}
\email{s.p.j.koenis@uu.nl}
\author{L. Maisel Licerán}
\author{H.T.C. Stoof}

\affiliation{%
Institute for Theoretical Physics and Center for Extreme Matter and Emergent Phenomena\\
Utrecht University, Princetonplein 5, 3584 CC Utrecht, The Netherlands
}%

\allowdisplaybreaks

\date{\today}

\begin{abstract}
    
    Conventional wisdom dictates that the conducting edge states of two-dimensional topological insulators of the \ce{Bi2Se3} family are protected by time-reversal symmetry.
    However, theoretical bulk calculations and a recent experiment show that the edge states persist in the presence of large external magnetic fields.
    To address this apparent contradiction, we have developed an analytical description for the edge-state wave function of a semi-infinite sample in a perpendicular magnetic field.
    Our description relies on the usual bulk Landau levels, together with additional states arising due to the presence of the hard wall, which are unnormalizable in the infinite system.
    The analytical wave functions agree extremely well with numerical calculations and can be used to directly analyze the behavior of the edge states in a magnetic field.
\end{abstract}

\maketitle


\section{\label{sec:Introduction}Introduction}

In recent years, interest in topological insulators (TIs) has surged due to the prospects of energy-efficient electronics and transport applications \cite{vzutic2004spintronics, hasan2011three, bernevig2013topological, pesin2012quantum, moore2010birth}.
These systems have gapped bulk states and gapless edge or surface modes, where the latter are topologically protected against disorder scattering.
A well-known example of this phenomenon is the quantum spin-Hall effect (QSHE), where two helical modes travel in opposite directions along the boundary of the system \cite{cage2012quantum}.

An example of such TIs is bismuth selenide (\ce{Bi2Se3}), a three-dimensional (3D) TI consisting of quintuple layers stacked on top of each other and bound by van der Waals forces.
It has attracted particularly significant attention due to its large band gap of $\SI{0.3}{\electronvolt}$ and its promising applicability \cite{xia2009observation, hsieh2009tunable, mazumder2021brief, aguilera2019many}.
Its topological properties arise from the spin--orbit coupling around the $\Gamma$ point \cite{qi2010quantum, yazyev2010spin}, which gives rise to topologically protected Dirac states at the top and bottom surfaces of thick samples.
Furthermore, in samples of less than six quintuple layers in thickness, the upper and lower surface modes couple via quantum tunneling and a thickness-dependent energy gap opens up \cite{chiatti20162d}.
For some thicknesses the gap is topological, causing a one-dimensional edge (1D) state to appear at the boundaries of the two-dimensional (2D) sample \cite{linder2009anomalous, maisel2024topology, hasan2010colloquium}.
These edge states have been studied in more detail theoretically \cite{liu2010oscillatory, lu2010massive} and also experimentally via angle-resolved photoemission spectroscopy \cite{zhang2010crossover, neupane2014observation} and scanning tunnelling microscopy (STM) \cite{moes2024characterization}.
The tunability of this energy gap makes these thin film TIs desirable for electronic device applications.
With these applications in mind, it is important to characterize and understand their edge states not only under ideal circumstances, but also in the presence of external perturbations.

It is often stated that the edge states of a TI such as \ce{Bi2Se3} are protected by time-reversal symmetry \cite{kane2005z,zhang2009topological,zhang2014robustness}.
However, a recent experiment shows with STM measurements that the edge states in \ce{Bi2Se3} nanosheets can exist when this symmetry is broken by a perpendicular magnetic field of up to at least $\SI{5}{\tesla}$ \cite{ThesisMoes2024}.
The effect of an out-of-plane magnetic field on the 2D surface states has been studied theoretically \cite{shafiei2022controlling,shafiei2024tuning}, and it has been found that the states remain topological for some thicknesses and field strengths. 
Theoretical calculations for the infinite Bernevig--Hughes--Zhang (BHZ) model for the QSHE have also found evidence that the 1D edge state can persist up to high magnetic fields \cite{zhang2014robustness}.
These calculations rely on the bulk-boundary correspondence to draw conclusions about the edge states \cite{mong2011edge}, but the latter are not computed explicitly.
Indeed, an analytic calculation of the edge states to investigate their behavior under a magnetic field has, to the best of our knowledge, not been performed.
The goal of this article is thus to provide a general description of the edge states of a \ce{Bi2Se3} nanosheet in a perpendicular magnetic field.
This sheds further light on the robustness of the edge state in the absence of time-reversal symmetry and lays the groundwork for future effective-model calculations.

Our article is structured as follows.
In Sec.\ \ref{sec: Bulk Hamiltonian} we outline the derivation of the Hamiltonian for a thin \ce{Bi2Se3} nanosheet, and its solution in the infinite-volume case. 
In Sec.\ \ref{sec:Edge Physics} we investigate the edge state of the \ce{Bi2Se3} nanosheet by introducing a hard wall in the system, effectively considering solutions in a semi-infinite nanosheet, and propose a basis of wave functions for analytically describing the 1D edge state at the boundary. Furthermore, we provide a step-by-step guide on how to find the analytical expression for the 1D edge mode traveling along the edge, yielding a dispersion relation of this edge state.
In Sec.\ \ref{sec:Results of the Model} we compare this dispersion relation to numerical calculations.
Finally, in Sec.\ \ref{sec:Outlook} we summarize the findings of this article, and discuss potential avenues for further research.

\section{Bulk Hamiltonian}
\label{sec: Bulk Hamiltonian}

For the characterization of the edge states of \ce{Bi2Se3}, it will be useful to first take a close look at the bulk states in a nanosheet of this material.
In this section we consider an infinite nanosheet of thickness $L_{z}$ lying parallel to the $xy$ plane.
The Hamiltonian for such a system can be derived from the 3D $\boldsymbol{k} \boldsymbol{\cdot} \boldsymbol{p}$ Hamiltonian for \ce{Bi2Se3} as explained in Refs.\ \cite{lu2010massive, zhou2008finite}, but we briefly outline the procedure here for completeness.

The starting point is the 3D Hamiltonian of the bulk material around the $\Gamma$-point, which for \ce{Bi2Se3} is given by the $\boldsymbol{k} \boldsymbol{\cdot} \boldsymbol{p}$ model \cite{liu2010model} 
\begin{equation}
\label{eq:3D_modelBi2Se3}
    H(\boldsymbol{k}, k_{z}) = \epsilon(\boldsymbol{k}) \mathbb{I}_4 + \begin{bmatrix}
    \mathcal{M}(\boldsymbol{k})& A_1 k_z & 0 & A_2 k_-\\
    A_1 k_z & -\mathcal{M}(\boldsymbol{k}) & A_2 k_- & 0\\
    0 & A_2 k_+ & \mathcal{M}(\boldsymbol{k}) & -A_1 k_z\\
    A_2 k_+ & 0 & -A_1 k_z & -\mathcal{M}(\boldsymbol{k})
    \end{bmatrix} .
\end{equation}
Here, $\boldsymbol{k} \equiv (k_{x}, k_{y})$, $k_{\pm} \equiv k_{x} \pm \iu k_{y}$, $\epsilon(\boldsymbol{k}) =  C - D_{1} k^{2}_{z} - D_2(k^{2}_{x} + k^{2}_{y})$, $\mathcal{M}(\boldsymbol{k}) = M - B_{1} k^{2}_{z} - B_{2}(k^{2}_{x} + k^{2}_{y})$, and $ \mathbb{I}_4$ is the $4 \times 4$ identity matrix. The values of the parameters are taken from Ref.\ \cite{zhang2009topological}.
This Hamiltonian is expressed in the basis \{${\vert \mathrm{Bi}^+, \uparrow \rangle},  {\vert \mathrm{Se}^-, \uparrow \rangle}, {\vert \mathrm{Bi}^+, \downarrow \rangle}, {\vert\mathrm{Se}^-, \downarrow \rangle}$\}, indexed by the orbitals closest to the Fermi surface and the spin of the electrons.

The effective Hamiltonian for an infinite nanosheet is obtained by restricting the size in the $z$ direction from $-L_{z}/2$ to $L_{z}/2$, thereby breaking the translational symmetry in this direction.
Hence, $k_z$ ceases to be a good quantum number, and we must make the substitution $k_z \rightarrow -\iu \partial_{z}$. This yields a Hamiltonian which can be solved at the 2D $\Gamma$ point $k_x = k_y = 0$.
The four states closest to the Fermi level are denoted by $\{\vert \mathrm{S}^{+}, \uparrow \rangle, \vert \mathrm{S}^{-}, \downarrow \rangle, \vert \mathrm{S}^{-}, \uparrow \rangle, \vert \mathrm{S}^{+}, \downarrow \rangle\}$, where $\mathrm{S}^{\pm}$ indicates that these are surface states with well-defined parity $\pm 1$.
Their expressions can be found in Ref.\ \cite{shan2010effective}, where they are denoted by $\{\vert \chi^{\uparrow} \rangle, \vert \varphi^{\downarrow} \rangle, \vert \varphi^{\uparrow} \rangle, \vert \chi^{\downarrow} \rangle\}$, respectively.
Note that the surface states $\mathrm{S}^{\pm}$ are not localized at only one of the top or bottom surfaces, but they are linear combinations of the latter.
In fact, as we explain below, in the limit of small thicknesses the surface states become coupled, thus the ``top/bottom'' index ceases to be a good quantum number.

We can obtain a low-energy effective model by projecting the Hamiltonian $H(\boldsymbol{k}, -\iu \partial_z)$ onto the subspace spanned by these four solutions, i.e.,
\begin{equation}
\label{eq:projection_into_subspace_hamiltonian}
    H_{\text{eff}}(\boldsymbol{k})_{\alpha \beta} = \int_{-L_{z}/2}^{L_{z}/2} \mathrm{d} z \, \Phi_{\alpha}^{\dagger}(z) H(\boldsymbol{k}, -\iu \partial_z)\Phi_{\beta}(z).
\end{equation}
Here, $\Phi_{\alpha}$ (with $\alpha = 1, \ldots, 4$) are the wave functions corresponding the four basis states above.
The Hamiltonian decouples into the block-diagonal form $H_{\text{eff}} = \operatorname{diag}(h'_+,h'_-)$, similar to the BHZ model, with  
\begin{equation}
    \label{eq:htz_no_magnetic_field}
    h'_{\tau_z}(\boldsymbol{k}) = E_0 - Dk^2 + \FermiVelocity (k_y \sigma_x - k_x \sigma_y) + \tau_z \left(\frac{\Delta}{2} - \Bparam k^2 \right) \sigma_z .
\end{equation}
Here, $\FermiVelocity$ is the Fermi velocity, $\Delta$ the (inverted) band gap energy at the $\Gamma$ point, $\sigma_{i}$ the Pauli matrices, and $\tau_z = \pm 1$ the so-called hyperbola index.
In this article, we will drop the irrelevant constant term $E_{0}$ from Eq.\ \eqref{eq:htz_no_magnetic_field}, and use units such that $\hbar=1$.

When the thickness $L_{z}$ is large, the four basis states $\Phi_{\alpha}$ are degenerate at the $\Gamma$ point and thus $\Delta = 0$.
The dispersion then consists of two gapless surface Dirac cones.
However, when the thickness is below six quintuple layers, a gap opens between these states and they remain only pairwise degenerate.
This is the regime of interest in this paper.
The two pairwise degenerate subspaces correspond to $\tau_{z} = \pm 1$.
This quantum number corresponds to the product of the parity of the $\Gamma$-point basis state times the spin projection, with the $\uparrow$ and $\downarrow$ spins associated with $+1$ and $-1$, respectively.
Thus, $\tau_{z} = +1$ for the $\{\vert \mathrm{S}^{+}, \uparrow \rangle, \vert \mathrm{S}^{-}, \downarrow \rangle\}$ subblock and $\tau_{z} = {-}1$ for the $\{{\vert \mathrm{S}^{-}, \uparrow \rangle}, {\vert \mathrm{S}^{+}, \downarrow \rangle\}}$ subblock.

Note that the Pauli matrices $\sigma_{i}$ are not directly the physical spin operators, as explained in Ref.\ \cite{maisel2024topology}.
Indeed, while they do couple the up and down spins, the latter are accompanied by the two different surface states $\mathrm{S}^{\pm}$ in each $\tau_{z}$ subblock.
Nevertheless, the time-reversal operator leaves the parities unaffected and thus still sends $\vec{\sigma} \rightarrow {-} \vec{\sigma}$ and $\vec{k} \rightarrow {-} \vec{k}$, so that the two subblocks are time-reversed copies of each other.

Next, we introduce a magnetic field in the $z$ direction, $\vec{B} = \BField  \hat{\vec{z}}$.
This adds a Zeeman term depending on the effective magnetic moment $\mu_{\text{eff}}$, and forces the replacement $\hat{p}_i \rightarrow \hat{p}_i + e A_i$, with $\vec{A}$ the magnetic vector potential and $e > 0$ the elementary charge.
We will later see that a convenient choice for the vector potential is $(0, \BField  x,0)$, so that $k_{y}$ remains well defined.
This yields the Hamiltonian which will be used throughout the rest of this article, namely
\begin{equation}
\label{eq:htz shifted}
    \begin{split}
        h_{\tau_z} &=  {-} D e\BField (\xi^2  -\partial_\xi^2 ) \mathbb{I}_{2} + \FermiVelocity \sqrt{e\BField }(\xi\sigma_x  + i \partial_\xi \sigma_y) \\
        &+\tau_z \bigg(\frac{\Delta}{2} -  \Bparam e\BField  (\xi^2  -\partial_\xi^2 ) \bigg)\sigma_z - \mu_\text{eff} \BField  \sigma_z .
    \end{split}
\end{equation}
For convenience, we have expressed everything in terms of the shifted position $\tilde{x} \equiv x + k_y \magneticLength^2$ and introduced the dimensionless parameter $\xi \equiv \tilde{x} / \magneticLength$, with $\magneticLength = 1 / \sqrt{e\BField }$ the magnetic length.

We note that since $\tau_z=\pm1$ represents which helical mode is under consideration, it is evident from the Hamiltonian in Eq.\ \eqref{eq:htz shifted} that the magnetic field does not couple the two modes together.

For all figures and explicit calculations in this paper we have taken $\Delta = \SI{0.07}{eV}$, $D = \SI{-0.08}{eV nm^2}$, $\Bparam= \SI{0.1}{eV nm^2}$, and $v_{\mathrm{F}} = \SI{0.295}{\electronvolt\nano\meter}$ , which describe nanosheets of four quintuple layers \cite{zhang2010crossover}.
These are not the values arising directly of the projection, but are fitted from experiment and most correctly reproduce the dispersion around the $\Gamma$ point when only four states are considered.
If an accurate description of the four-band subspace closest to the Fermi level is desired in terms of projected parameters only, then at least eight bands must be used in the projection \cite{maisel2024topology}.
As this is analytically much more cumbersome, we employ the fitted four-band model instead.
We take $\mu_{\text{eff}} = g \mu_{\mathrm{B}}$, with $\mu_{\mathrm{B}}$ the Bohr magneton and $g$ the electronic $g$ factor whose value we take as 2 for simplicity and lack of experiments.

Note that $h_{\tau_z}$ is invariant under the simultaneous transformations $\tau_z \rightarrow -\tau_z$, $\BField  \rightarrow -\BField $, and $k_y \rightarrow -k_y$, together with a reordering of the basis that switches the spins. Hence, without loss of generality, we assume that $\BField  > 0$ in the rest of the article.

\subsection{\label{subsec:Solution Bulk}Bulk Solution with \texorpdfstring{$\BField  \neq 0$}{B!=0}}

We will now solve the eigenvalue equation arising from Eq.\ \eqref{eq:htz shifted} to characterize the bulk solutions of this system.
This has been done in Ref.\ \cite{zhang2014robustness}, but we outline the calculation here as it will be useful for Sec.\ \ref{sec:Edge Physics}. 
The problem is equivalent to a system of two coupled harmonic oscillators and can be solved analytically by introducing the operators
\begin{subequations}
\label{eqs:ladder operators}
    \begin{align}
        \hat{a}^{\dagger} &= \frac{1}{\sqrt{2}}(\xi - \partial_\xi),\\
        \hat{a} &= \frac{1}{\sqrt{2}}(\xi + \partial_\xi),
    \end{align}
\end{subequations}
where we call $\hat{a}^{\dagger}$ the creation or raising operator and $\hat{a}$ the annihilation or lowering operator. 
Using these, the Hamiltonian $h_{\tau_z}$ is rewritten as
\begin{widetext}
    \begin{equation}
    \label{eq:htz ladder expression}
       h_{\tau_z} = \begin{bmatrix}
             \tau_z \frac{\Delta}{2}-\mu_{\text{eff}} \BField  - 2\sqrt{e \BField }(D+\tau_z B)(\hat{a}^{\dagger}\hat{a} + \frac{1}{2}) & \FermiVelocity\sqrt{2 e \BField }\hat{a}\\
            \FermiVelocity\sqrt{2 e \BField }\hat{a}^{\dagger} &  - \tau_z \frac{\Delta}{2}+\mu_{\text{eff}} \BField  - 2\sqrt{e \BField }(D-\tau_z B)(\hat{a}^{\dagger}\hat{a} + \frac{1}{2})
        \end{bmatrix}.
    \end{equation}
\end{widetext}
Each diagonal term depends on $\hat{a}^{\dagger}\hat{a} + \frac{1}{2}$ and hence corresponds to a regular harmonic oscillator.
The eigenfunctions of this term alone are the normalized Hermite functions \cite{rushka2020completely}
\begin{equation}
\label{eq: eigenfunction diagonal H}
    \phi_n(x,k_y) =\frac{1}{\sqrt{2^n n! \magneticLength\sqrt{\pi}}} H_n(\xi) \ec^{-\xi^{2}/2},
\end{equation}
with $H_n(\xi)$ the Hermite polynomial of order $n$ in terms of $\xi$.
Equation \eqref{eq: eigenfunction diagonal H} gives the eigenfunctions of the diagonal part of the Hamiltonian of Eq.\ \eqref{eq:htz ladder expression}, and the eigenfunctions to the full Hamiltonian can be constructed via the Ansatz
\begin{align}
\label{eq:excited state h_tz shifted}
    \psi^\pm_{\tau_zn,\text{bulk}}(x,k_y) &=
    \begin{bmatrix}
    \alpha^\pm_{\tau_zn} \phi_{n-1}(x,k_y)\\
    \beta^\pm_{\tau_zn} \phi_{n}(x,k_y)
    \end{bmatrix} \\
    &= \frac{1}{\sqrt{2^n n! \magneticLength\sqrt{\pi}}} \ec^{-\xi^{2}/2} \begin{bmatrix}
        \sqrt{2 n} \alpha^\pm_{\tau_zn} H_{n-1}(\xi) \\
         \beta^\pm_{\tau_zn} H_n(\xi) \nonumber
    \end{bmatrix}
\end{align}
for some pair of numbers $v^\pm_{\tau_zn} = [\alpha^{\pm}_{\tau_zn}, \beta^{\pm}_{\tau_zn}]^{\mathsf{T}}$ with $n \geq 1$. 
We call wave functions of this form Landau states (LSs).
The superscript $\pm$ represents two possible inequivalent solutions for a given $\tau_z$ and $n$.
Letting the Hamiltonian act on this Ansatz and solving for nontrivial $v^\pm_{\tau_zn}$ yields the Landau levels (LLs) of the system, which read
\begin{equation}
    \label{eq:Landau Energies}
    E_{\tau_zn}^{\pm} = -2 e\BField Dn + e\BField  \Bparam \tau_z 
    \pm \textstyle{\frac{1}{2}} \sqrt{{\cal A}_{\tau_zn}} ,
\end{equation}
with
\begin{equation}
    {\cal A}_{\tau_zn} = \Big( 2 e\BField  \left( D-\Bparam n\tau_z \right) - 2\mu_{\text{eff}} \BField  +\Delta \tau_z\Big)^2
        + 8e\BField n \FermiVelocity^2 .
\end{equation}
The corresponding (unnormalized) eigenstate coefficients are
\begin{equation}
\label{eq:eigenvector bulk}
    v_{\tau_zn}^{\pm} = 
    \begin{bmatrix}
        \Delta \tau_z+ 2e\BField  \left(D-2 \Bparam n \tau_z \right)-2\mu \BField  \pm \sqrt{{\cal A}_{\tau_zn}} \\
     2 \FermiVelocity \sqrt{2e\BField n} 
    \end{bmatrix} .
\end{equation}
We then have an infinite set of states with $n \ge 1$.
Importantly, however, there is one additional solution for each $\tau_z$, namely $n=0$, with energy $E^{-}_{{+}1, 0}$ ($E^{+}_{{-}1, 0}$) for $\tau_{z} = {+}1$ ($\tau_{z} = {-}1$).
Notice that in this case the superscript is fully determined by the sign of $\tau_z$, and as such, we will drop it in the rest of this article for $n=0$.
The wave functions in both cases read $\psi_{0,\text{bulk}}(x,k_y) = [0, \phi_{0}(x,k_y)]^{\mathsf{T}}$, and hence $ v_{\tau_zn} = [0,1]^\mathsf{T}$.
The energy levels of the bulk states for $\tau_z=+1$ correspond to the dashed grey lines of Fig.\ \ref{fig:Dispersion_edge_state_B1}.

It is worth mentioning here that the wave function of Eq.\ \eqref{eq:excited state h_tz shifted} with $n<0$ would in principle also be a solution of the Hamiltonian, with an eigenvalue given by Eq.\ \eqref{eq:Landau Energies}. 
However, as $\phi_n$ diverges to infinity for negative $x$, these solutions are not normalizable and are therefore not present in the bulk system. 
Nevertheless, we will see that they play a crucial role in the description of the edge state, and return to this point in a later section.
For reference, the energy of these non-normalizable Landau states is shown by the dotted green lines in Fig.\ \ref{fig:Dispersion_edge_state_B1}.

\section{\label{sec:Edge Physics}Edge Physics}

We now turn to the primary focus of this article: the edge state of the \BiSe nanosheets in a magnetic field.
We introduce the edge state by considering a semi-infinite \ce{Bi2Se3} nanosheet of thickness $L_{z}$ lying parallel to the $xy$ plane as in the last section, but now only in the region $x \ge 0$.
The edge states are localized close to the boundary at $x = 0$ and their momentum $k_{y}$ along the $y$ axis is a good quantum number. 
We assume a hard-wall boundary condition at $x=0$. 
We choose $\vec{A} = (0, \mathcal{B}x, 0)$, so that the translational symmetry is only broken in the direction perpendicular to the edge and thus the Hamiltonian describing the system is exactly the same as in Sec.\ \ref{sec: Bulk Hamiltonian}.
Therefore, finding an expression for the edge state amounts to solving the eigenvalue equation arising from the Hamiltonian in Eq.\ \eqref{eq:htz shifted} with the added boundary condition $\psi(x=0) = 0$.
To the best of our knowledge, this eigenvalue problem with $\BField \neq 0$ cannot be solved analytically for arbitrary $k_y$.
As such, in the rest of this article we will concern ourselves with finding approximate solutions to the equation. We do this by using an approximate basis of wave functions at each value of $k_y$ and solving the Schrödinger equation within this basis.
While this basis is not exactly orthogonal, we will nevertheless find excellent agreement with the numerical results.

\subsection{\label{subsec:basis for wave function} Basis for the edge-state wave function}
\begin{figure}
	\centering
	\includegraphics[width=1\linewidth]{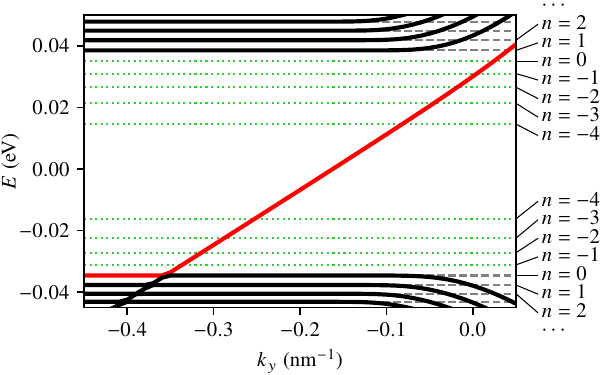}
	\caption{Numerical calculation of the spectrum of the system for $\BField  = \SI{1}{\tesla}$ and $\tau_z={+}1$.
		For clarity, $\tau_z=-1$ is not shown.
		The dashed lines are the LLs given by Eq.\ \eqref{eq:Landau Energies} for different values of $n$.
		The black lines represent LLs which correspond to a physical bulk state, and the green lines LLs with states that are not present in bulk.
		The red line corresponds to the edge state, as this is the band that forms the Dirac cone with its $\tau_z = -1$ counterpart (not shown here).
		It can be seen that every band acquires the bulk LL energy when $k_{y} \rightarrow {-}\infty$, and tends towards the dashed black lines.}
	\label{fig:Dispersion_edge_state_B1}
\end{figure}

Before we give the basis of wave functions for the edge state, it is insightful to first explain the general idea behind the construction of these wave functions. 
In the previous section we have found that the Landau states of Eq.\ \eqref{eq:excited state h_tz shifted} are solutions to the Schrödinger problem with the Hamiltonian of Eq.\ \eqref{eq:htz shifted}. 
The only problem with applying these solutions to the edge state is, of course, that they do not adhere to the hard-wall boundary condition at $x=0$. 
However, we now show that we can make a modification to the Landau-state wave functions such that they adhere to the boundary conditions while still approximately solving the Schrödinger equation with remarkable accuracy. 
We call these modified states the Dirichlet Landau states (D-LSs), and the energy corresponding to these solutions is taken to be equal to the energy of Eq.\ \eqref{eq:Landau Energies} for the original LS. 
Looking at Fig.\ \ref{fig:Dispersion_edge_state_B1}, we see that the edge state dispersion (shown in red) at multiple $k_y$'s approaches or crosses the Landau energies (dashed and dotted lines). 
The central idea of this paper is then as follows: when the wave number $k_y$ is such that the energy of the edge state is close to a Landau energy, we can accurately describe the edge state wave function by the D-LS corresponding to this Landau energy. 
This gives us a list of specific wave numbers $k_y$ for which we can accurately obtain the edge state. We then treat these solutions, the D-LSs, as a basis for solving the entire dispersion relation of the edge state.

For reasons that will be obvious later, it is convenient to split the D-LS basis into three distinct classes of wave functions, namely 
\begin{enumerate}
    \item Dirichlet bulk states (D-bulk),
    \item Dirichlet non-normalizable Landau states (D-NLS),
    \item Dirichlet complex Landau states (D-CLS).
\end{enumerate}
As we will discuss later, the energy of the D-NLSs and D-CLS is calculated with Eq.\ \eqref{eq:Landau Energies} with $n\leq0$. 
The energy levels of the D-NLS are shown by the green dotted lines in Fig.\ \ref{fig:Dispersion_edge_state_B1}.
For each class we give a brief interpretation and some physical justification in the following sections. 
It is worth mentioning that the construction of the D-LSs was initially motivated by the observation that the D-bulk states were remarkably accurate approximations to the numerical solution of the edge state wave function at specific $k_y$'s.
This invited a more careful analysis of the problem, resulting in the construction of the D-NLS and D-CLS, which turned out to yield accurate approximations of the edge state as well.
For a more detailed discussion and a derivation of the three classes, the reader is referred to the appendices where applicable.  

The general form of the basis wave functions of each class is given by 
\begin{equation}
\label{eq:general_wave_function_edge}
    \psi_{\tau_zn}^\pm = \mathcal{N}_{\tau_zn}^\pm \left(1- \ec^{-\Lambda_-(E_{\tau_zn}^\pm) x} \right) \begin{bmatrix}
        \alpha_{\tau_zn}^\pm \phi_{n-1}(x,k_y) \\
        \beta_{\tau_zn}^\pm \phi_n(x,k_y)
    \end{bmatrix},
\end{equation}
where $\mathcal{N}_{\tau_zn}^\pm$ is a normalization constant. The energy $E_{\tau_z,n}^\pm$ is given by Eq.\ \eqref{eq:Landau Energies}, with different choices for $n$ and the corresponding branch (that is, either $E^+$ or $E^-$) for each class of wave functions. 
We will see that for some classes we either pick the $+$ or $-$ solution, and as such, we drop the superscript in those cases for clarity. 
The only difference between the classes lies in their choice of solution branch $\pm$, $n, \alpha$, and $\beta$, as we explain in the following sections.

It can be seen that the sole difference between the D-LSs and the ``regular'' Landau states is the factor $(1- \ec^{-\Lambda_-(E_{\tau_zn,i}) x})$. 
This is the modification we alluded to at the start of this section, and forces the hard-wall boundary condition to be satisfied.
In this factor, we have used the exponential decay parameter
\begin{equation}
\label{eq:lamda}
    \Lambda_-(E) = \sqrt{\frac{2D(E + k_y^2D) - \FermiVelocity^2 - F(\Delta - 2B k_y^2)- \sqrt{R(E)}}{2(D^2 - F^2)}},
\end{equation}
where $R(E) = 4E^2F^2 + \FermiVelocity^4 - 2F \Delta \FermiVelocity^2 + D^2\Delta^2 - 4 ED(\FermiVelocity^2 - F\Delta)$.  The derivation of this prefactor relies on the calculation of the edge state for $\BField=$ \SI{0}{\tesla}, which is explained in detail in Appendix\ \ref{sec:SolutionB0}. 
In short, it can intuitively be viewed as a factor that dominates the behavior of the wave function near the edge, but changes it only marginally for $x\gg 0$.

\subsubsection{Dirichlet bulk states}
\label{subsubsec:Bulk state}
We refer to the first class of the basis wave functions as the Dirichlet bulk states. This name reflects that they arise directly from the bulk-state calculation in Sec.\ \ref{sec: Bulk Hamiltonian}, with a modification to adhere to the hard-wall (Dirichlet) boundary condition. As such, these wave functions have the form of Eq.\ \eqref{eq:general_wave_function_edge} where the parameters $\alpha$ and $\beta$ are given by 
\begin{equation}
\label{eq:basis_D-LS}
    \begin{bmatrix}
    \alpha_{\tau_zn,\text{D-bulk}} \\
    \beta_{\tau_zn,\text{D-bulk}}
    \end{bmatrix} =  v_{\tau_zn}
\end{equation}
as in Eq.\ \eqref{eq:eigenvector bulk}, where $n>0$. We remind that $ v_{\tau_z, 0} = [0,1]^\mathsf{T}$ for both $\tau_z=\pm1$. As we have seen in Sec.\ \ref{sec: Bulk Hamiltonian}, the bulk wave functions have two solution branches $v_{\tau_zn}^\pm$ for $n>0$. 
This raises an obvious question, namely which branch to use for the D-bulk states.
As we will discuss later in this section, we can deduce by physical arguments that for $k_y \ll 0$ the two branches of the edge state must approach the Landau energies $E_{\tau_z, 0}$. 
By noting that the two $\tau_z$ cases must form a Dirac cone, we can deduce whether each branch goes upwards or downwards. 
Alternatively, and perhaps more insightfully, one can solve the edge state numerically and take note of the general shape of the dispersion relation. 
This is done in Fig.\ \ref{fig:Dispersion_edge_state_B1} for $\tau_z=1$ and $\BField=$\SI{1}{\tesla}. 
In this figure the edge state is shown in red, and we see that it crosses all the Landau energies of the $v_{\tau_zn}^+$ branches.
Therefore, for $\tau_z=+1$ and $n>0$ we use $v_{\tau_zn}^+$ and $
E_{\tau_zn}^+$ in Eqs.\ \eqref{eq:general_wave_function_edge} and \eqref{eq:basis_D-LS}. 
Conversely, for $\tau_z = -1$ we use $v_{\tau_zn}^-$ and $E_{\tau_zn}^-$. Since the superscript is fully determined by the sign of $\tau_z$, we drop it for clarity. 

As discussed before, we state that the wave function $ \psi_{\tau_zn,\text{D-bulk}}$ from Eq.\ \eqref{eq:general_wave_function_edge} is a good approximation to the true edge state for the precise $k_y$ when its energy is equal to $E_{\tau_zn}$, that is, the $k_y$ at which the red edge state crosses the dashed Landau levels in Fig.\ \ref{fig:Dispersion_edge_state_B1}.
We will denote the wave number uniquely attributed to $E_{\tau_zn}$ ($n>0$) as $k_{y,\tau_zn}$. 
A method of finding the set $\{k_{y,\tau_zn} \}$ via a self-consistency relation can be found in Appendix\ \ref{sec:wave functions at discrete ky}. 
In brief, the wave number can be calculated from the fact that if one takes the expectation value of the Hamiltonian for these wave functions and sets it equal to a Landau energy, the wave number $k_y$ is the only unknown in the equation.
Note that this method does not rely in any way on solving the differential equation numerically.
In this section we also show that Eq.\ \eqref{eq:basis_D-LS} is a remarkably accurate approximation of the edge state wave function at $k_{y,\tau_zn}$, and thus our choice of this basis vector is justified.

An additional comment needs to be made about the wave function $ \psi_{\tau_z,0,\text{D-bulk}}$, i.e., the case $n=0$. As can be seen in Fig.\ \ref{fig:Dispersion_edge_state_B1}, the edge state never actually crosses $E_{\tau_z,0}$, but only approaches it asymptotically. 
To explain this behavior, it is insightful to consider the coordinate transformation $\xi = x/l_\BField + k_y l_\BField$ that was performed to obtain the solution of the bulk Hamiltonian. 
We know that the Hermite function $\phi_{n}(x,k_y)$ as defined by Eq.\ \eqref{eq: eigenfunction diagonal H} is localized around $\xi=0$, and thus around $x = - k_y/l_\BField^2$. 
Therefore, for $k_y \ll 0$, the wave function is localized at $x \gg 0$. 
Since the Hermite functions have Gaussian decay, the wave function essentially already satisfies the Dirichlet boundary condition in this case.
In other words, we expect the wave function of the edge state to be identical to the solution of the bulk Hamiltonian in the regime $k_y \ll 0$, with an identical energy.
Looking at Eq.\ \eqref{eq:general_wave_function_edge}, it can readily be verified that for these values of $k_y$, we indeed have $\psi_{\tau_z,\text{D-bulk}}(x,k_y) = \psi_{\tau_z,0,\text{bulk}}(x,k_y)$ as the exponential term is approximately zero.
The asymptotic behavior of the edge state can then be attributed to the decaying influence of the edge at $x=0$, which is zero in the limit $k_y \rightarrow-\infty$. 
Note that this implies that, as opposed to the case $n>0$, we cannot attribute a single $k_{y,\tau_z,0}$ to the energy $E_{\tau_z,0}$. Instead, we expect the wave function $ \psi_{\tau_z0,\text{D-bulk}}$ to be a good approximation to the edge state over a large range of $k_y \ll0$.
 
\subsubsection{Dirichlet non-normalizable Landau states}
\label{subsec:Dirichlet non-normalizable Landau states}
We now draw attention to the Hermite functions in Eq.\ \eqref{eq: eigenfunction diagonal H}. 
As mentioned previously, the Hermite functions with $n\geq0$ are not the only possible solutions to the diagonal components of the Hamiltonian in Eq.\ \eqref{eq:htz ladder expression}.
Additionally, there are non-normalizable solutions which go to infinity for $x \rightarrow -\infty$.
We call these functions the non-normalizable Hermite functions $\phi_n(x,k_y)$, with $n < 0$.
In the usual treatment of the quantum harmonic oscillator these functions are ignored, because they constitute non-normalizable and thus unphysical solutions. 
However, in the case of the edge state, these functions exist in the domain $[0,\infty)$, and are therefore normalizable.
Hence, it is not unreasonable to suggest that they could manifest themselves in the functional form of the edge-state wave function for some $k_y$.

This observation gives rise to our second class of basis functions, which we call the Dirichlet non-normalizable Landau states (D-NLSs).
All of these wave functions have at least one component which contains a negative order Hermite function, and have a real energy which is once again calculated using Eq.\ \eqref{eq:Landau Energies}.
The amount of D-NLSs one has to include in the analysis depends on the precise parameters of the system and the magnetic field strength. 
For example, as is also shown in Fig.\ref{fig:Dispersion_edge_state_B1}, for the parameters under consideration in this article, Eq.\ \eqref{eq:Landau Energies} yields real values for all $n \geq -4$ ($n \geq -5$) for $\tau_z = 1$ ($\tau_z = -1$). 
Each value of $n\leq0$ corresponds to two D-NLSs, at energies $E_{\tau_z,n}^\pm$, with the exception of $n=0$, which only has a single D-NLS at $E_{1,0}^+$ ($E_{-1,0}^-$) for $\tau_z=1$ ($\tau_z=-1$), as the other branch is the regular bulk state $\psi_{\tau_z0,\text{bulk}}$.
Since for $n<0$ both branches are used for different wave functions in this class, the superscript is not determined by $\tau_z$, and we cannot remove it, as opposed to the D-bulk wave functions.

To calculate the prefactors $\alpha_{\tau_z n}^\pm$ and $\beta_{\tau_z n}^\pm$ of the D-NLS in Eq.\ \eqref{eq:general_wave_function_edge}, we do a similar calculation as in Sec. \ref{subsec:Solution Bulk}. 
We construct the ansatz $[\alpha^{\pm}_{\tau_zn} \phi_{n-1}, \beta^{\pm}_{\tau_zn} \phi_n]^{\mathsf{T}}$, and let the Hamiltonian from Eq.\ \eqref{eq:htz ladder expression} act on it. 
Solving the system for nontrivial  $\alpha_{\tau_z n}^\pm$ and $\beta_{\tau_z n}^\pm$ yields the desired prefactors.
The only difference with the previous calculation is that for the D-CLS we in general have $n \leq 0$, so $\phi_n$ is no longer the usual Hermite function of Eq.\ \eqref{eq: eigenfunction diagonal H}.
Instead, we introduce the negative order Hermite functions, where the function for $n=-1$ is given by
\begin{equation}
    \phi_{-1}(x,k_y) = \frac{2}{\sqrt{ \magneticLength\sqrt{\pi}\operatorname{ln} 2}} H_{-1}(\xi) \ec^{-\xi^2/2}.
\end{equation}
In this equation, we note that $H_{-1}(x) = \frac{1}{2}\sqrt{\pi} \ec^{x^2}\text{erfc}(x)$, with $\operatorname{erfc}(x)$ the complementary error function. The prefactor is chosen such that the function is normalized with respect to $\xi$ in the domain $[0,\infty)$.

Hermite functions of lower order $n$ can be obtained via the relation
\begin{equation}
	\hat{a} \phi_n(x) = d_n\phi_{n-1}(x),
\end{equation}
where $d_n$ is a known constant. 
A full treatment of the extension of the harmonic oscillator to the regime of $n<0$ is given in Appendix \ref{sec:negative order hermite functions}, together with a table for the constant $d_n$.
Furthermore, in this appendix we describe how the ladder operators from Eqs.\ \eqref{eqs:ladder operators} act on the non-normalizable Hermite functions, which is used to calculate the prefactors  $\alpha_{\tau_z n}^\pm$ and $\beta_{\tau_z n}^\pm$. 
To the best of our knowledge, this treatment and application of the non-normalizable harmonic oscillator functions is a novel ingredient of our work.

Just like the functions from the first class, the wave functions in this class are good solutions when the energy of the edge state is equal to the corresponding energy of the underlying Landau state, in this case $E_{\tau_z,n}^\pm$ with $n \leq 0$.
In practice this means that the wave functions of Eq.\ \eqref{eq:basis_D-LS} are accurate solutions at discrete values for the wave number, which we will call $k_{y, \tau_z,\text{D-NLS}}^\pm$, and these values are found by the same procedure as in the previous section.
Examples of some D-NLS wave functions can be found in App.\ \ref{sec:wave functions at discrete ky}.

\subsubsection{Dirichlet complex Landau states}
\label{subsec:Dirichlet complex Landau states}
In addition to the D-NLSs, it is natural to look at Hermite functions of even lower order. 
This poses a problem, however, because the energy as calculated by by Eq.\ \eqref{eq:Landau Energies} turns out to be complex for low enough order.
For the \BiSe parameter values at $\BField =  \SI{1}{\tesla}$, this is the case for $n<-4$ ($n<-5$) for $\tau_z=1$ ($\tau_z=-1$).
Additionally, the eigenvectors are also complex, with $v_{\tau_zn}^+$ the complex conjugate of $v_{\tau_zn}^-$ in these cases.
Hence, we call these states the Dirichlet complex Landau states.
We interpret this as these states not having a well-defined energy, but being spread out around the real part of this energy. We will briefly return to this point in App.\ \ref{sec:negative order hermite functions}.

Since $E_n^+$ is the complex conjugate of $E_n^-$ for all complex values of the energy, i.e., the real part of the energy is degenerate, we can make a superposition of two solutions giving a state with a real energy and real wave function. In contrast to the wave functions of other classes, which precisely follow Eq.\ \eqref{eq:general_wave_function_edge}, we thus have that the wave functions of the third class follow

\begin{align}
    \label{eq:wave function D-CL}
    \psi_{\tau_zn,\text{D-CLS}}(x)& =\\
    \mathcal{N}_{n,\text{D-CLS}}\bigg( &\big(1 - \ec^{-\Lambda_-(E^+_{\tau_zn,\text{D-CLS}}) x} \big)\begin{bmatrix}
        \alpha_{\tau_zn}^+ \phi_{n-1}(x,k_y) \\
        \beta_{\tau_zn}^+ \phi_n(x,k_y)
    \end{bmatrix}\nonumber \\  + \, &\big(1 - \ec^{-\Lambda_-(E^-_{\tau_zn,\text{D-CLS}}) x} \big)\begin{bmatrix} 
        \alpha_{\tau_zn}^- \phi_{n-1}(x,k_y) \\
        \beta_{\tau_zn}^- \phi_{n}(x,k_y) 
    \end{bmatrix}\bigg), \nonumber
\end{align}
which is simply a superposition of two copies of the original wave-function form, one for each sign. 
Notice that this wave function is real for all D-CLSs.

To determine the precise value of $k_y$ at which the D-CLS is valid, which we call $k_{y,\tau_z,n,\text{D-CLS}}$, we follow the same procedure as for the other classes, as outlined in App.\ \ref{sec:wave functions at discrete ky}.
However, since the Landau energy corresponding to this state is complex, we only take the real part of the energy into account for this calculation.
We will see that this provides good results.

It is important to note that, in theory, one could introduce any number of D-CLSs for the analysis of the edge state. 
In this article however, we only take the highest-order D-CLS into account, as we have found that this is sufficient to achieve a good edge state wave function.
In App. \ref{sec:lower_order_D-CLS} we briefly discuss the effect of adding more D-CLSs, and see that this yields overwhelmingly diminishing returns after just one D-CLS.

\subsection{\label{subsec:Interpolating} Basis expansion}

\begin{figure*}[ht]
	\includegraphics[width=1\linewidth]{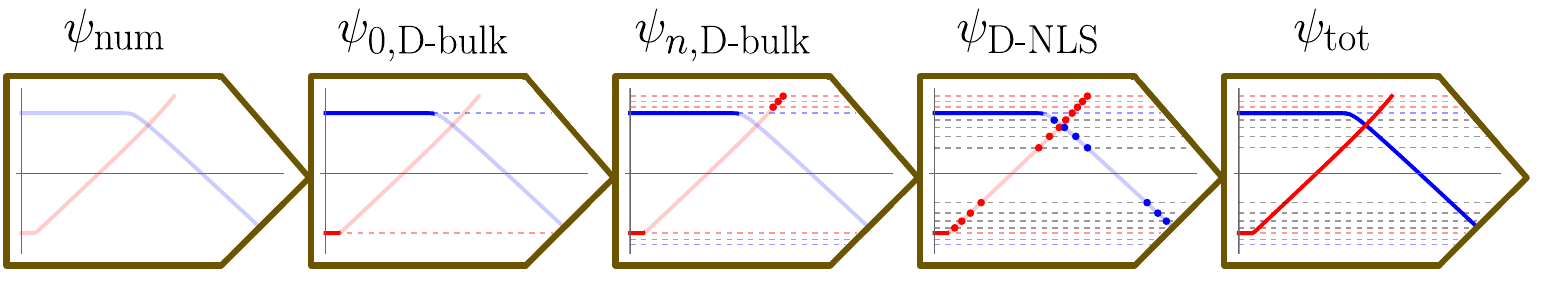}
	\caption{Graphical summary of this work, a piecewise construction of an analytical function for the edge state.
		Each panel shows a dispersion relation of the one-dimensional \BiSe edge state with the wave number $k_y$ on the $x$ axis.
		The red (blue) color represents the $\tau_z =+1$ ($\tau_z = -1$) edge state.
		In the first panel we show the numerical dispersion relation in faded colors, with wave function $\psi_{\text{num}}$.
		Full colours represent the fact that we know the analytical wave function corresponding to this $k_y$. 
		Note that we use the numerical solution only for reference, as it is not used in the analytical calculation. 
		In the second and third panel we add $\psi_{n, \text{D-bulk}}$ with $n=0$ and $n>0$ respectively. The dashed lines correspond to the Landau energy. In the fourth panel we add the D-NLSs at their corresponding $k_y$. The D-CLS is not shown here. Finally, in the last panel we interpolate between these known basis vectors as per Sec.\ \ref{subsec:Interpolating} to obtain the full dispersion relation, achieving the goal of this article.}
	\label{fig:Graphical_Summary}
\end{figure*}

Using the states discussed in Sec.\ \ref{subsec:basis for wave function}, we would now like to express the total wave function of the edge state in this basis. 
This is straightforwardly done by making a linear superposition of the basis vectors for each possible value of $k_y$.
The wave function at a given $k_y$ then looks like

\begin{equation}
\label{eq: psi sum general}
	\psi_{\tau_z,\text{tot}}(x,k_y) = \sum_{i \in I_2} c_i(k_y) \psi_{\tau_z,i}(x,k_{y,i})
\end{equation}
where $I$ is a set of indices which reflect the basis states that one chooses to include at this specific $k_y$.
Furthermore, $\psi_{\tau_z,i}$ is an element of the set of basis states as defined in Sec.\ \ref{subsec:basis for wave function}, and $k_{y,i}$ the corresponding discrete value of the wave number associated with $\psi_{\tau_z,i}$.
As discussed before, for $\psi_{0,\tau_z,\text{D-bulk}}$ this discrete value of $k_y$ does not exist, so then we will simply take the input $k_y$ for this basis wave function.
We stress that this is the only wave function out of the entire basis set that has an explicit $k_y$-dependence. 
When $\psi_{0,\tau_z,\text{D-bulk}}$ is not part of the basis set for $\psi_{\tau_z,\text{tot}}(x,k_y)$, all $k_y$-dependence of the full solution stems from the coefficients $c_i(k_y)$.

One may wonder if we always need to include the D-NLSs and D-CLS, since an accurate description of the bulk wave functions does not need them at all. 
We argue that the addition of these wave functions is necessary for any topological edge state.
In terms of parameter values, this means that for the magnetic fields considered in this article we must have $B \cdot \Delta>0$ \cite{zhang2014robustness}, as this causes the Hamiltonian of Eq.\ \eqref{eq:htz shifted} to be topological. 
The reasoning is as follows: we have seen that the edge state at a certain energy is most accurately described by a wave function whose underlying bulk state has that same energy. However, we know that the (real part of the) energy of the D-NLSs and D-CLS lies in the gap of the dispersion relation. 
When a conducting edge state exists, it by definition passes through this gap, and thus crosses these energies. 
Therefore, if an edge state exists in the system, the D-NLSs and D-CLS must be included for an accurate description of its wave function.

To calculate the coefficients $c_i$ in Eq.\ \eqref{eq: psi sum general} for a given $k_y$ we impose that $\psi_{\tau_z,\text{tot}}(x,k_y)$ be an eigenvector of $h_{\tau_z}$ with eigenvalue $E(k_y)$, and that the expectation value of $h_{\tau_z}$ with respect of $\psi_{\tau_z,\text{tot}}(x,k_y)$ be equal to that same $E(k_y)$.
This yields a linear system of equations, which together with the normalization condition determine $E$ and all $c_i$.
These equations are the time-independent Schrödinger equations expressed in a non-orthogonal basis and are given by
\begin{equation}
    \label{eq:self consistency alpha}
    \sum_j\langle \psi_i |h_{\tau_z}| \psi_j \rangle c_j = \sum_j E \langle \psi_i | \psi_j \rangle c_j ,
\end{equation}
where index $j$ represents the components of the total wave function as in Eq.\ \eqref{eq: psi sum general}.
Note that the matrix elements on the right-hand side of Eq.\ \eqref{eq:self consistency alpha} are generally nontrivial because our basis states $\psi_i$ are in general not fully orthogonal.
Once these $c$'s have been obtained for a single $k_y$, one has an analytical approximation of the edge state at that wave number without relying on the numerical solution, thereby achieving our goal.

\subsection{Recap of method}
\label{subsec:Recap}
In this section we take a step back and summarize the step-by-step recipe to arrive at the full edge-state dispersion relation, which is the core result of this paper.
Given $\tau_z$ and a value of $\BField  > 0$, the steps are as follows:
\begin{enumerate}[topsep=0pt,itemsep=-1ex,partopsep=1ex,parsep=1ex]
  \item Calculate the bulk eigenstates of Eq.\ \eqref{eq:excited state h_tz shifted}, and construct the LL wave functions $\psi^\pm_{\tau_zn,\text{bulk}}(x,k_y)$. 
  \item Calculate the LL energies $E_{\tau_zn}$ up to some $n$ using Eq.\ \eqref{eq:Landau Energies}.
  Use these energies to construct $\psi_{\tau_zn \text{,D-bulk}}$ with Eq.\ \eqref{eq:basis_D-LS} for these same $n$.
  As discussed in Appendix\ \ref{sec:wave functions at discrete ky}, calculate the discrete values of $k_y$ that correspond to these functions via a self-consistency relation.
  \item Calculate the D-NLSs according to Sec.\ \ref{subsec:Dirichlet non-normalizable Landau states}, and find the appropriate $k_{y, \tau_zn,\text{D-NLS}}$.
  The amount of D-NLSs one has to include depends on the precise system parameters and value of $\BField$. 
  Similarly, calculate the highest order D-CLS as described in Sec.\ \ref{subsec:Dirichlet complex Landau states}.
  Together with step 3, this  yields a list of solutions to the edge-state Hamiltonian at the points $(k_{y,\tau_z,i}, E_{\tau_z,i})$, with $\psi_{\tau_z,i}$ the corresponding solution. Sort this list such that $k_{y,\tau_z,i}$ is increasing in value.
  \item Pick a $k_y$ at which the edge-state wave function is to be computed.
  Propose a linear combination of the basis set of wave functions as in Eq.\ \eqref{eq: psi sum general}, and calculate the coefficients using Eq.\ \eqref{eq:self consistency alpha}.
  \item Finally, repeat step 4 for many $k_y$ to obtain the full dispersion relation.
\end{enumerate}
A graphical summary of the recap, represented as the piecewise construction of the dispersion relation, is shown in Fig.\ \ref{fig:Graphical_Summary}.
Regarding step 4, we remind that one has the freedom to choose how many wave functions are taken into account for the linear combination of Eq.\ \eqref{eq: psi sum general}.
In general, we have found that for a given $k_y$, it yields good results to include into Eq.\ \eqref{eq: psi sum general} just the two wave functions whose $k_{y,\tau_z,i}$ is closest to the value of $k_y$.
However, as $\psi_{\tau_z,0,\text{D-bulk}}$ does not have this discrete value of $k_y$, we take this solution into account when the edge state approaches the bulk energy with $n=0$, i.e. when $k_y$ is lower than the lowest $k_{y,\tau_z,i}$.
To summarize, we recommend the following:
\begin{itemize}[topsep=0pt,itemsep=-1ex,partopsep=1ex,parsep=1ex]
\item For $k_y < k^\mp_{y,\tau_z,-1,\text{D-NLS}}$ with $\tau_z = \pm 1$, include $ \psi_{\tau_z,0,\text{D-bulk}}(k_y)$ and $\psi^\mp_{\tau_z,-1,\text{D-NLS}}$, where we remind that we only have explicit $k_y$-dependence in the former basis function.
\item For $k_y \geq k^\mp_{y,\tau_z,-1,\text{D-NLS}}$ with $\tau_z = \pm 1$, include $\psi_{\tau_z,i}$ and $\psi_{\tau_z,i+1}$ such that $k_{y,\tau_z,i} < k_y  <k_{y,\tau_z,i+1} $. We remind that these wave functions can be D-CLSs, D-NLSs or D-bulk ($n>0$) states, and that none of them have explicit $k_y$-dependence.
\end{itemize}
We have found that this convention yields excellent results, and we will use it in the rest of the article. 

\begin{figure}[!t]
    \centering
    \includegraphics[width=1.0\linewidth]{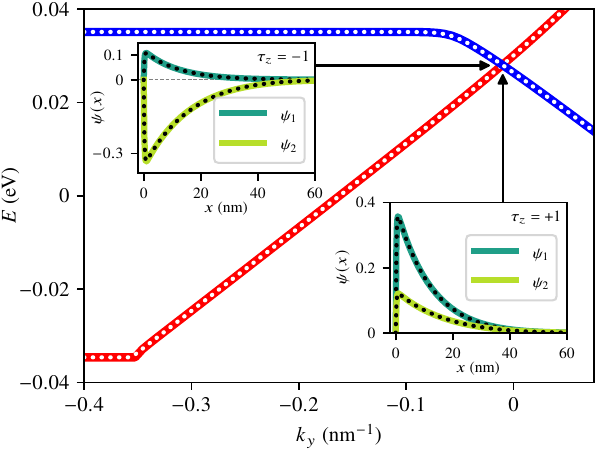}
    \caption{Dispersion relation of the edge state at $\BField = \SI{1}{\tesla}$.
    The white dots are the numerical solutions while the full lines are obtained via the method described in Sec.\ \ref{subsec:Recap}.
    The red and blue lines correspond to $\tau_z = {+}1$ and $\tau_z =-1$, respectively.
    It can be seen that our method agrees extremely well with the numerical solution.
    The insets show the two wave functions of the bands at the gamma point indicated by the black arrow, with $\psi_1$ and $\psi_2$ the first and second component of the wave function, respectively.
    The black dots show the numerical calculation and the solid lines correspond to our approximation.}
    \label{fig:Dispersion_Edge_state_a_priori_B1}
\end{figure}

\section{\label{sec:Results of the Model}Results of the model}

In this section we first compare the method as described in the last section to the numerical solution of the differential equation.
The case we discuss is the edge state with $\BField = \SI{1}{\tesla}$.
At the end of this section we use the analytical method to obtain experimentally observable STM spectra.

\begin{figure}[!t]
    \centering
    \includegraphics[width=1.0\linewidth]{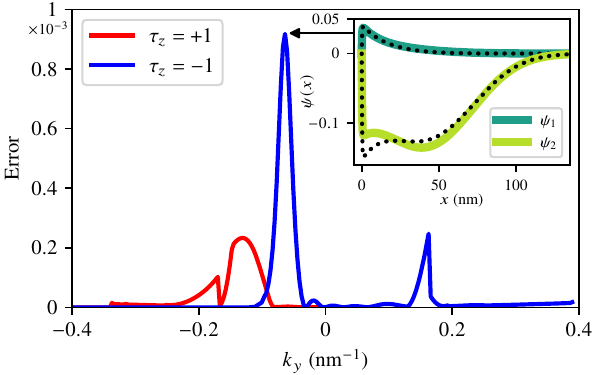}
    \caption{$L^{2}$ difference between the numerical solution and our analytical approximation, i.e., $\int_0^\infty \mathrm{d}x \, \vert\hspace{-0.4mm}\vert\psi_{\text{num}}(x) - \psi(x)\vert\hspace{-0.4mm}\vert^2$.
    The red and blue lines correspond to $\tau_z = {+}1$ and $\tau_z =-1$, respectively.
    We draw attention to the fact that the error is almost zero at the gamma point ($k_y = -0.01$) (cf. Fig.\ \ref{fig:Dispersion_Edge_state_a_priori_B1}).
    The inset shows the comparison at the attached highest-error point and reveals that even in this case the approximation is relatively accurate.}
    \label{fig:psi_error_B1}
\end{figure}

In Fig.\ \ref{fig:Dispersion_Edge_state_a_priori_B1} we show the dispersion relation obtained via this method.
The dotted lines correspond to numerical simulations, while the full lines correspond to the wave function obtained via our analytical method. 
The dispersion relations match almost perfectly, and in fact the error never exceeds $\SI{0.15}{\milli\electronvolt}$.
Fig.\ \ref{fig:psi_error_B1} shows the total difference between wave functions of the two methods for all values of $k_y$.
This error is calculated as $\int_0^\infty \mathrm{d}x \, \vert\hspace{-0.4mm}\vert\psi_{\text{num}}(x) - \psi(x)\vert\hspace{-0.4mm}\vert^2$ and is seen to be very small everywhere.
The inset of Fig.\ \ref{fig:psi_error_B1} shows the worst-performing wave function for $\BField = \SI{1}{\tesla}$, which is at $k_y \approx \SI{-0.06}{\nano\meter\tothe{-1}}$ and $\tau_z = {-}1$.
Even at this point, the approximation is very close to the numerical solution, with an $L^2$ error below $0.1\%$.
Therefore, we conclude that our method yields a good approximation of the exact solution for both the energy and wave function at this magnetic field.
We have found similar results for all $\BField \in [0.1,10] \; \si{\tesla}$.

Fig.\ \ref{fig:Dispersion_Edge_state_a_priori_B1} also shows that the two $\tau_z$ branches behave asymmetrically.
This is due to the broken particle--hole symmetry in the system, which manifests in the Hamiltonian of Eq.\ \eqref{eq:htz shifted} via to the addition of a diagonal momentum-dependent term.
The nonzero $D$ effectively pushes the Dirac point upwards in energy, causing the dispersion relation of the edge state to become asymmetric around $E=0$.

\begin{figure}[ht]
    \centering
    \includegraphics[width=1.0\linewidth]{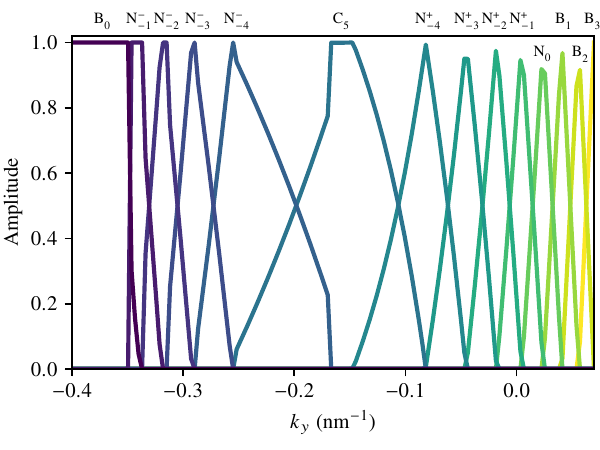}
    \caption{Parameters $c_i$ from Eq.\ \eqref{eq: psi sum general} for the wave functions with $B=1$ T and $\tau_z =1$. Each colour represents a different part of the wave function. In this figure we represent the D-bulk states as $\mathrm{D}_n$, the D-NLSs as $\mathrm{N}^\pm_n$ and the D-CLSs as $\mathrm{C}_n$. Notice that since this basis is in general not exactly orthonormal, the squared factors at a single $k_y$ do not necessarily exactly add up to one. It can be seen that at many wave numbers the wave function consists of just a single component, indicating that a single basis vector is sufficient to describe the edge state at those wave numbers.}
    \label{fig:Blending_parameters_t1_B1}
\end{figure}

Fig.\ \ref{fig:Blending_parameters_t1_B1} shows the evolution of the coefficients $c_i$ introduced in Eq.\ \eqref{eq: psi sum general} as a function of $k_y$ for $\BField = \SI{1}{\tesla}$ and $\tau_z = {+}1$.
Each color represents the $c_i$ corresponding to a different wave function $\psi_{\tau_z,i}$.
We see that for wave numbers between $\SI{-0.35}{\nano\meter\tothe{-1}}$ and $0$ the wave function is fully composed of (a combination of) the D-CLS and D-NLSs, practically without any error in a subset of this domain (see also Fig.\ \ref{fig:psi_error_B1}).
This is conclusive evidence that these basis vectors, and thus the negative order Hermite functions, cannot be ignored in the treatment of this system.

\begin{figure}[ht]
	\centering
	\includegraphics[width=1\linewidth]{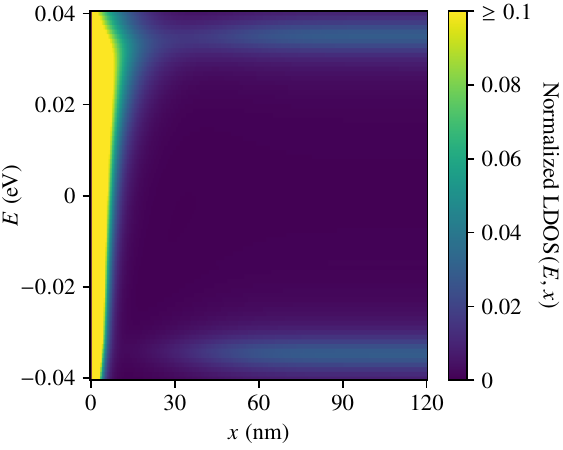}
	\caption{Calculation of the local density of states (LDOS) for the edge state in \BiSe via our method in a perpendicular magnetic field of $\SI{1}{\tesla}$. The LDOS measurement is modeled as a Cauchy (Lorentzian) distribution with $\gamma = \SI{7.5}{meV}$. In addition to the edge state, we have included the D-bulk states of order $n=0$ of each $\tau_z$ for this calculation. They can be clearly identified at $E=\pm \SI{35}{meV}$ eV.}
	\label{fig:LDOS_B1}
\end{figure}

Finally, we show in Fig.\ \ref{fig:LDOS_B1} the local density of states (LDOS) of the edge state in a $\SI{1}{\tesla}$ magnetic field, as observable in STM measurements. 
Readily identifiable are the two bulk energies, around $\pm0.035$ eV, which correspond to the edge state at very negative wave numbers.
By comparing the LDOS of this article with experimentally obtained data, one could verify the validity of our analytic approximation.
\section{\label{sec:Outlook}Conclusion and outlook}

In this article we have derived an analytical description of the edge-state wave functions of a \BiSe nanosheet under a perpendicular magnetic field as a function of the wave number $k_{y}$.
The approximation is based on the observation that the edge state wave function inherits the form of the Landau states when the energy is equal to the corresponding Landau energy.
Furthermore, we have introduced the D-NLS and D-CLS, which are usually non-normalizable bulk-like states appearing solely due to the presence of the edge, and are constructed from negative order Hermite functions.
We have shown that they are crucial for describing the edge state in the energy range corresponding to the bulk band gap.
Using these states, we have devised an analytical description of the wave function for general $k_y$.
This yields accurate results that agree with numerical calculations.

Our analytical description enables subsequent calculations that can shed light on the nature of the edge state and its behavior under external fields and perturbations.
For example, one can derive an effective Hamiltonian for the nanosheet edge state in a perpendicular magnetic field that can be used for a detailed study of the edge physics alone and as a starting point for the detailed study of impurity scattering.
Furthermore, we can investigate the behavior of the Dirac point of the system for high values of the magnetic field.
Our calculation shows that time-reversal symmetry is not necessary to explain the robustness of the edge state.
We argue that even though time-reversal symmetry is broken, the perpendicular magnetic field does not couple the two helical modes to each other.
As such, scattering between them is impossible and the edge state remains robust.
One possible area of further research is to consider a magnetic field with an in-plane component.
This magnetic field would couple the two modes together, presumably extinguishing the edge state, similar to the effect of magnetic impurities \cite{ThesisMoes2024}.
We hope that our work stimulates more experimental work on these exciting features of topological edge states.

\section*{Acknowledgments}

We thank Daniel Vanmaekelbergh and Ingmar Swart for useful discussions about the experiments.
This work is supported by the research program ``Materials for the Quantum Age'' (QuMat).
This program (registration number 024.005.006) is part of the Gravitation program financed by the Dutch Ministry of Education, Culture and Science (OCW).


\appendix

\section{\label{sec:SolutionB0}Edge-state solution for \texorpdfstring{$\BField = 0$}{B=0}}
In this section we find the wave function of the edge state without a magnetic field. 
For the Hamiltonian of Eq.\ \eqref{eq:htz_no_magnetic_field} one can plug in the ansatz $\psi(x) = \vec{v} \ec^{-\Lambda x}$, with $\vec{v}$ a two-component vector. 
This yields four solutions with nontrivial $\vec{v}$, out of which two are normalizable. 
The full solution is a superposition given by
\begin{equation}
    \label{eq:solution edge state}
    \psi_{\BField=0} (x) = A \vec{v}_{\Lambda_-}\ec^{-\Lambda_-(E) x} +  B \boldsymbol{v}_{\Lambda_+}\ec^{-\Lambda_+(E) x}.
\end{equation}
The $\Lambda$'s are given by
\begin{equation}
    \Lambda_\pm(E) = \sqrt{\frac{2D(E + k_y^2D) - \FermiVelocity^2 \pm F(\Delta - 2B k_y^2)\pm \sqrt{R(E)}}{2(D^2 - F^2)}},
\end{equation}
where  $R(E) = 4E^2F^2 + \FermiVelocity^4 - 2F \Delta \FermiVelocity^2 + D^2\Delta^2 - 4 ED(\FermiVelocity^2 - F\Delta)$. The eigenvectors are given by
\begin{equation}
    \label{eq:eigenvector B=0}
    \vec{v}_{\Lambda_\pm} = \begin{pmatrix}
        \tau_z(\Delta/2 + B \Lambda_\pm^2) + E - D \Lambda_\pm^2 \\
        v_F\Lambda_\pm
    \end{pmatrix}.
\end{equation}
The energy $E$ and prefactors $A$ and $B$ can be numerically obtained from the boundary condition $\psi(0) = 0$ and normalization. 
Hence, in the $\BField=0$ case the solution is a sum of two decaying exponentials.
It can be verified that for the parameter values of \BiSe (Sec.\ \ref{sec: Bulk Hamiltonian}) we find $\Lambda_->\Lambda_+$.
Hence, the behavior around $x=0$ is governed by $\Lambda_-$, i.e. $\psi_{\BField=0} (x \approx 0) = \tilde{\boldsymbol{v}} (\ec^{-\Lambda_- x} - 1)$, where $\tilde{\boldsymbol{v}} = A \vec{v}_+ = B \vec{v}_-.$ 
This has an important implication for the wave function of the edge state with nonzero magnetic field.
One may notice that the magnetic field almost does not change the Hamiltonian if we were looking very close to the edge, i.e., for $x \approx 0$.
The only difference is the term $\mu_e \BField\sigma_z$, which we can absorb in parameter $\Delta$, yielding an effective band gap $\Delta_{\text{eff}} =\Delta - 2\mu_{\text{eff}}\BField $.
For all the magnetic field strengths considered in this article, we have $\BField \ll \Delta/2\mu_{\text{eff}}$, which means that we can safely take $\Delta_{\text{eff}} \approx \Delta$ for all further calculations.
Therefore, we argue that close to the edge, we can ignore the effect of the magnetic field, and we retrieve a wave function that resembles the $\BField = 0$ case.
As a consequence, if we have a solution of the Hamiltonian with magnetic field that does not adhere to the boundary condition at $x=0$, we can multiply this solution with the factor $(1-  \ec^{-\Lambda_-(E) x})$ to presumably get a very close approximation of the true edge state at energy $E$.
This insight is how we constructed Eq.\ \eqref{eq:general_wave_function_edge}, and we will see in the following sections that it indeed provides remarkably good solutions to the Hamiltonian.

Of course, one may ask what happens if the magnetic field is of such strength that the band gap energy is comparable to the Zeeman splitting, i.e. $\BField = \Delta/ 2\mu_{\text{eff}}$, as in this case we are no longer allowed to ignore the effect of the magnetic field near the edge.
For the parameter values of \BiSenospace, this is the case for a magnetic field of the order of $300$ T. 
Not only are measurements at such magnetic fields far beyond the capabilities of current experimental setups, it has been shown by Zang et al. \cite{zhang2014robustness} that the topological edge state persists only up to a critical magnetic field of $\BField_{\text{crit}}=\hbar \Delta/2eB$. 
As $\BField_{\text{crit}}= 230$ T for the parameters used in this article, we expect that the approximation $\Delta_{\text{eff}} =\Delta$ is adequate for all observable edge states in \BiSenospace.
	
	\section{Numerical solution of the Hamiltonian}
Throughout this article, we have compared the analytical approximation of the edge state with the numerical solution of the Hamiltonian of Eq.\ \eqref{eq:htz shifted} to test the validity of the approximation. In this section, we briefly discuss how to obtain this numerical solution.

We start by noting that the eigenfunctions of the Hamiltonian are wave functions in the semi-infinite domain. Hence, to find these functions numerically, we need perform a coordinate transformation such that we can properly discretize this domain. We chose to transform the $x$-coordinate for nonzero magnetic field to $\zeta$ using

\begin{equation}
	\label{eq:zeta definition}
	\zeta = \frac{2}{\pi}\text{arctan}( x/l_\BField),
\end{equation}
where we remind that the magnetic length $l_\BField = 1/\sqrt{e \BField}$.
From this expression it is obvious that $\zeta \in [0,1)$, which is a domain that can be discretized via a finite grid.
With this coordinate transformation, it can be shown that the Hamiltonian of Eq.\ \eqref{eq:htz shifted} can be rewritten as

\begin{eqnarray}
	\label{eq:htz_numerical}
	&&h_{\tau_z}(k_y) = \tau_z\frac{\Delta}{2}\sigma_z - \mu_{\text{eff}} \BField \sigma_z \\
	&&-\frac{D}{l_\BField^2} \mathcal{D}(\zeta,k_y) \mathbb{I}_{2}  - \frac{B}{l_\BField^2} \tau_z\mathcal{D}(\zeta,k_y)\sigma_z \nonumber\\
	&&	+\frac{v_F}{l_\BField} \left[\left(\text{tan}(\pi \zeta/2) + k_y l_\BField\right)\sigma_x  -i \frac{2}{\pi} \text{cos}^2(\pi \zeta/2) \partial_\zeta \sigma_y\right],\nonumber\\
\end{eqnarray}
with 
\begin{eqnarray}
	\mathcal{D}(\zeta,k_y) &=& \left(\text{tan}(\pi \zeta/2) + k_y l_\BField \right)^2 \\ &&+\frac{4}{\pi^2} \text{cos}^4(\pi \zeta/2) \left( \text{tan}(\frac{\pi}{2} \zeta)\partial_\zeta - \partial_\zeta^2  \right).\nonumber
\end{eqnarray}
After imposing the boundary conditions $\psi(\zeta=0)=0$ and $\psi(\zeta=1)=0$, this Hamiltonian equation can be solved by a finite element solver of choice. 
The solution in $\zeta$-space is then converted to real space by inverting Eq.\ \eqref{eq:zeta definition}, yielding the numerical wave functions and dispersion relations. 
For all figures involving the numerical solution in this article, we have used the function NDEigensystem in Mathematica, which uses Arnoldi iteration to solve the differential equation, with a maximum cell measure of $10^{-4}$.

\section{Negative-order Hermite functions}
\label{sec:negative order hermite functions}
We have seen in the main text that to accurately describe the edge state, we need the Hermite functions with order $n<0$. In this section, we give a derivation for these functions, and we see under which conditions they solve the Hamiltonian of Eq.\ \eqref{eq:htz shifted}. 

\subsection{Negative-order Hermite polynomials}
To derive the Hermite polynomials with negative index, our first instinct may be to look at the recurrence relation for the polynomials, which is given by 
\begin{equation}
\label{eq:recurrence hermite}
H_{n+1}(x) = 2xH_{n}(x) - 2n H_{n-1}(x).
\end{equation}
However, it is obvious that we are unable to calculate $H_{-1}(x)$ from $H_{0}(x)$ and $H_{1}(x)$, since this function gets cancelled for $n=0$. Luckily, there is another recurrence relation in integral form given by 
\begin{equation}
    \int^x_0\ec^{-y^2}H_n(y) \, \text{d}y = H_{n-1}(0) - \ec^{-x^2}H_{n-1}(x),
\end{equation}
which is not as troublesome \cite{NIST:DLMF18.17.E4}. Plugging in $n=0$ yields
\begin{equation*}
    \int^x_0\ec^{-y^2} \, \text{d}y = H_{-1}(0) - \ec^{-x^2}H_{-1}(x),
\end{equation*}
where we recognise the integral as $\frac{\sqrt{\pi}}{2}\text{erf}(x)$, with $\operatorname{erf}(x)$ the error function. We now make the substitution $H_{-1}(x) = \ec^{x^2} f(x)$, which after some rearrangements yields
\begin{equation*}
    f(x)- f(0) = - \frac{\sqrt{\pi}}{2}\text{erf}(x).
\end{equation*}
This is solved by
\begin{equation*}
    f(x) = \frac{\sqrt{\pi}}{2}(1-\text{erf}(x)) = \frac{\sqrt{\pi}}{2}\text{erfc}(x),
\end{equation*}
where $\operatorname{erfc}(x) = 1 - \operatorname{erf}(x)$ is the complementary error function. This gives us as a final result for the Hermite polynomial of order $n=-1$, namely
\begin{equation}
\label{eq:H-1appendix}
    H_{-1}(x) = \frac{\sqrt{\pi}}{2}\ec^{x^2} \text{erfc}(x).
\end{equation}
As a sanity check, we can verify that this function solves the characteristic differential equation for Hermite polynomials 
\begin{equation}
    \frac{\text{d}^2H_n(x)}{\text{d}x^2} - 2 x \frac{\text{d}H_n(x)}{\text{d}x} + 2 n H_n(x) =0
\end{equation}for $n=-1$.

With $H_{-1}(x)$ determined, we can easily obtain all other Hermite polynomials with negative index with the help of Eq.\ \eqref{eq:recurrence hermite}. 
The first few are given by
\begin{subequations}
\label{eq:negative hermite polynomials}
    \begin{align}
        \label{eq:negative hermite polynomial H-1}
        H_{-1}(x) &= \frac{1}{2}\sqrt{\pi} \ec^{x^2}\text{erfc}(x),\\
        \label{eq:negative hermite polynomial H-2}
        H_{-2}(x) &= \frac{1}{2}\Big(1-\sqrt{\pi} \ec^{x^2} x\; \text{erfc}(x)\Big),\\
        \label{eq:negative hermite polynomial H-3}
        H_{-3}(x) &= \frac{1}{8}\Big(\sqrt{\pi} (2x^2 + 1) \ec^{x^2}\text{erfc}(x) -2x\Big).
    \end{align}
\end{subequations}
While these functions are not polynomials, to avoid confusion we group them with the positive index functions and call them all Hermite polynomials.

\subsection{Non-normalizable bulk solutions}
We know that the Hermite functions for $n\geq0$ have the shape $\phi_n(x) \propto H_n(x)\ec^{-x^{2}/2}$, so it is reasonable to suggest that this could also be true for $n<0$.
With the Hermite polynomials from last section in hand, we see that on the domain $(-\infty,\infty)$ these functions are non-normalizable for all $n<0$, which is why they are not included in the usual treatment of the regular harmonic oscillator. 
However, we are studying the domain $[0,\infty)$, and in this case, these functions are normalizable.
Therefore, we cannot rule out that they play a role in the semi-infinite system. 

We will now prove that $H_n(x)\ec^{-x^{2}/2}$ with $n<0$ indeed solves the diagonal part of the Hamiltonian of Eq.\ \eqref{eq:htz shifted}. For ease of notation we will assume $k_y=0$ and $l_\BField=1$. The final results are easily converted to solutions with general $k_y$ and $l_\BField$.
Since we are looking at the extensions of the bulk solutions, we do not impose any boundary conditions except normalizability over $x$ from $0$ to $\infty$. 
From Sec. \ref{subsec:Solution Bulk}, we know that if we have a solution $\phi_n(x)$ to ($\hat{a}^{\dagger}\hat{a} + \frac{1}{2}$), then $\hat{a}\phi_n(x) \propto \phi_{n-1}$ is another solution. 
We remind that $\hat{a}^{\dagger}$ and $\hat{a}$ are the creation and annihilation operators, respectively.
However, since by definition $\hat{a}\phi_0 = 0$, we cannot directly obtain $\phi_{-1}(x)$ from the positive $n$ solutions. Therefore, we will be inspired by Eq.\ \eqref{eq: eigenfunction diagonal H} to assume $\phi_{-1}(x) = N_{-1} H_{-1}(x)\ec^{-x^2 / 2}$, and show that it is an eigenfunction of the differential equation
\begin{equation}
    \frac{1}{2}\left(x^2 - \frac{\text{d}^2}{\text{d}x^2}\right)\phi_{-1}(x) = \lambda_{-1}\phi_{-1}(x).
\end{equation}
Keeping in mind that
\begin{equation*}
    \frac{\text{d}}{\text{d}x}\text{erfc}(x) = \frac{2}{\sqrt{\pi}}\ec^{-x^2},
\end{equation*}
it is easy to show that this is indeed the case, and that the function has an eigenvalue of $\lambda_{-1} = -\frac{1}{2}$. 
We can calculate the normalisation factor $N_{-1}$ by integrating this function over $x$ from $0$ to $\infty$, after which we conclude

\begin{equation}
\label{eq: psi n-1}
     \phi_{-1}(x) = \frac{2}{\sqrt{\operatorname{ln} (2) \sqrt{\pi}}} H_{-1}(x)\ec^{-\frac{x^2}{2}}.\\
\end{equation}

After getting this foothold in the realm of the negative $n$, we can construct all the other functions using
\begin{equation}
    \label{eq:general negative hermite functions}
    \phi_{n}(x) = A_n (\hat{a}_{-})^{-n-1}\phi_{-1}(x) \quad \forall \, n<0,
\end{equation}
where $A_n$ is a normalisation factor dependent on $n$, as of yet only known for $n=-1$. From the commutation relation of the ladder operators and Eq.\ \eqref{eq:recurrence hermite} we can deduce that $\phi_n(x) \propto H_n(x)\ec^{-x^{2}/2}$ is indeed a solution for negative $n$ as well, with only the prefactor $A_n$ to be determined. In theory, if we normalise each wave function by hand, Eq.\ \eqref{eq:general negative hermite functions} is enough to obtain everything we need to construct the D-NLS and D-CLS. However, it is more insightful to take a closer look at how the ladder operators behave with these negative order Hermite functions on the domain $[0,\infty)$.

\subsection{\label{subsubsec:Ladder operators}Ladder operators on the semi-infinite domain}
We know that the ladder operators act on a state $\phi_n(x)$ such that 
$\hat{a}_\pm \phi_n(x) \propto \phi_{n\pm 1}(x)$, but for negative $n$ we do not know yet what the prefactor for this operation is. We define
\begin{subequations}
    \begin{align}
        \label{eqs:ladder operators cn}
        \hat{a}^{\dagger}  \phi_n(x) &= c_n\phi_{n+1}(x), \\
        \label{eqs:ladder operators dn}
        \hat{a} \phi_n(x) &= d_n\phi_{n-1}(x)
    \end{align}
\end{subequations}
and we want to find an expression for $c_n$ and $d_n$ for negative $n$.
Of course, $c_n= \sqrt{n+1}$ and $d_n= \sqrt{n}$ for $n\geq0$. 
We start by considering the integral 
\begin{equation}
\label{eq:normalisation psin-1}
    \int_0^\infty (\hat{a}\phi_n(x))^*(\hat{a}\phi_n(x)) \,\text{d}x = |d_n|^2, 
\end{equation}
where we have used Eq.\ \eqref{eqs:ladder operators dn} and the knowledge that $\phi_{n-1}$ should be normalised on $[0,\infty)$. Furthermore, using Eq.\ \eqref{eqs:ladder operators}, we can partially integrate Eq.\ \eqref{eq:normalisation psin-1} to obtain

\begin{equation}
\label{eq:partial integration psin-1}
   |d_n|^2 = \frac{1}{\sqrt{2}}[(\hat{a}\phi_n(x))^*\phi_n(x)]_0^\infty + \int_0^\infty (\hat{a}^{\dagger}\hat{a} \phi_n)^*\phi_n \,\text{d}x.
\end{equation}
It is important to draw our attention for a moment to the boundary term in Eq.\ \eqref{eq:partial integration psin-1}. Notice that it cancels for positive $n$ for both the $(-\infty,\infty)$ and $[0,\infty)$ domain. This is because $\phi_n(x)$ must be normalisable for all $n$, so it must vanish at $\pm \infty$. Furthermore, for $n>0$ either $\phi_n(x)$ or $\phi_{n-1}(x)$ is an odd function, and hence the product must be zero at $x=0$. However, as we will later see, this is \emph{not} the case for $n<0$, and thus the boundary term cannot be ignored.

To continue the calculation, we identify 
\begin{equation}
\label{eq:number operator n<0}
    \hat{a}^{\dagger}\hat{a} \phi_n = n \phi_n ,
\end{equation}
which we can deduce from expression \eqref{eq:general negative hermite functions} and the commutation relation $[\hat{a},\hat{a}^{\dagger}]=1$, to arrive at the final expression
\begin{equation*}
    |d_n|^2 =n - \frac{1}{\sqrt{2}}[(\hat{a}\phi_n(x))^*\phi_n(x)]\bigg\rvert_{x = 0}.
\end{equation*}
Doing a similar calculation for $c_n$ yields
\begin{equation*}
    |c_n|^2 =n + 1 + \frac{1}{\sqrt{2}}[(\hat{a}^{\dagger}\phi_n(x))^*\phi_n(x)]\bigg\rvert_{x = 0}.
\end{equation*}
Notice that this result is consistent with the ladder operators for $n \geq 0$. 
We still have the freedom of choosing a phase factor for $c_n$ and $d_n$. 
The only restriction for this choice is that it must be consistent with Eq.\ \eqref{eq:number operator n<0}, and thus we enforce $c_{n-1}d_n = n $. 
A suitable choice for $n=-1$ is
\begin{subequations}
    \begin{align}
        \label{eqs:prefactors ladder operators cn -1}
        c_{-1} &= \sqrt{2} \sqrt{ \frac{1}{\sqrt{2}}[(\hat{a}^{\dagger}\phi_{-1}(x))^*\phi_{-1}(x)]\bigg\rvert_{x = 0}},\\
        \label{eqs:prefactors ladder operators dn -1}
        d_{-1} &= -\sqrt{-1 - \frac{1}{\sqrt{2}}[(\hat{a}\phi_{-1}(x))^*\phi_{-1}(x)]\bigg\rvert_{x = 0}},
    \end{align}
\end{subequations}
and for $n\neq-1$
\begin{subequations}
\label{eqs:prefactors ladder operators cn dn}
    \begin{align}
        \label{eqs:prefactors ladder operators cn}
        c_n &= \sqrt{n + 1 + \frac{1}{\sqrt{2}}[(\hat{a}^{\dagger}\phi_n(x))^*\phi_n(x)]\bigg\rvert_{x = 0}},\\
        \label{eqs:prefactors ladder operators dn}
        d_n &= \operatorname{sgn}(n) \sqrt{n - \frac{1}{\sqrt{2}}[(\hat{a}\phi_n(x))^*\phi_n(x)]\bigg\rvert_{x = 0}},
    \end{align}
\end{subequations}
where we have used the sign function instead of a minus sign to make the expression consistent with $n\geq0$.
These prefactors ensure normalization for the wave functions, and can be used to calculate all wave functions according to Eq.\ \eqref{eq:general negative hermite functions}. 
After reintroducing $k_y$ and $l_\BField$, the first few Hermite functions are given by
\begin{subequations}
\label{eq:negative index hermite functions}
    \begin{align}
        \label{eq:negative index hermite function -1}
        \phi_{-1}(x) &= \frac{2}{\sqrt{\operatorname{ln}(2) l_\BField\sqrt{\pi}}} H_{-1}(\xi) \ec^{-\xi^{2}/2},\\
        \label{eq:negative index hermite function -2}
        \phi_{-2}(x) &= \frac{2\sqrt{2}}{\sqrt{(1-\operatorname{ln}2)l_\BField\sqrt{\pi}}} H_{-2}(\xi) \ec^{-\xi^{2}/2},\\
        \label{eq:negative index hermite function -3}
        \phi_{-3}(x) &= \frac{8}{\sqrt{(2\operatorname{ln}2 -1 )l_\BField\sqrt{\pi}}} H_{-3}(\xi) \ec^{-\xi^{2}/2}, 
    \end{align}
\end{subequations}
where we remind that $\xi = x/l_\BField + k_yl_\BField$, and all the wave functions are normalized from $\xi=0$ to $\xi=\infty$.

A comment needs to be made about the case separation for $n =-1$. 
The separation stems from the fact that in Sec. \ref{subsec:Solution Bulk} we defined $\phi_0(x)$ as normalized over ${(-\infty,\infty)}$, while $\phi_{-1}(x)$ is normalized over $(0,\infty)$. 
To address this mismatch in definition, $c_{-1}$ needs to have an extra factor $\sqrt{2}$, to make sure that Eqs. \eqref{eqs:prefactors ladder operators cn dn}, \eqref{eq:negative index hermite functions}, and \eqref{eq: eigenfunction diagonal H} are all consistent with this difference in normalization range. 
Alternatively, we could have avoided this mismatch by imposing that the Hermite functions from Eq.\ \eqref{eq: eigenfunction diagonal H} must be normalized from $\xi=0$ to $\xi = \infty$.
This would have effectively only changed the numerical value of $\mathcal{N}_{\tau_zn}^\pm$ in Eq.\ \eqref{eq:general_wave_function_edge}.
However, for the sake of consistency with the regular harmonic oscillator, we choose not to do this, and instead solve the issue by simply changing $c_{-1}$. Note that this only plays a role in the calculation of $[\alpha_{\tau_z,\text{D-NLS}}, \beta_{\tau_z,\text{D-NLS}}]^\mathsf{T}$ in Sec.\ \ref{subsec:Dirichlet non-normalizable Landau states}, as this calculation uses an Ansatz with both $\phi_0$ and $\phi_{-1}$ and thus $c_{-1}$ enters the equation.

The first few values of $c_n$ and $d_n$ are shown in Table\ \ref{tab:prefactors negative ladders}. 
For $n < 0$ these prefactors seem to follow the pattern
\begin{subequations}
\label{eq:prefactors direct calculation}
    \begin{align}
        \label{eq:prefactors direct calculation cn}
        c_n &= \sqrt{1 + \delta_{n, -1}}\sqrt{n+1 + \frac{(-1)^n}{\mathcal{C}_n-\operatorname{ln}(2)}},\\
        \label{eq:prefactors direct calculation dn}
        d_n &= {-}\sqrt{n + \frac{(-1)^n}{\mathcal{C}_n-\operatorname{ln}(2)}},
    \end{align}
\end{subequations}
with
\begin{equation}
    \mathcal{C}_n = \sum\limits_{i=1}^{-n-1}\frac{(-1)^{i+1}}{i},
\end{equation}
where we note that $C_0 = 0$.
We have found that Eqs.\ \eqref{eq:prefactors direct calculation} are correct to \textit{at least} $n =-50$, but we have not proven the general case.

With the explicit values of $c_n$ and $d_n$ in hand, we can now diagonalize the bulk Hamiltonian of Eq.\ \eqref{eq:htz ladder expression} with the Ansatz of Eq.\ \eqref{eq:excited state h_tz shifted}, where now $n < 1$ is allowed.

\begin{table}[ht]
\centering
\caption{Prefactors of the ladder operators for negative index.}

\begin{tabular}{l|l|l}
$n$  & $c_n$ & $d_n$ \\ \hline
$-1$ & $\sqrt{\frac{2}{\operatorname{ln}2}}$ & $-\sqrt{\frac{1}{\operatorname{ln}2}-1}$  \\
$-2$ & $\sqrt{\operatorname{ln}2-1}$  & $-\sqrt{\frac{1}{1-\operatorname{ln}2}-2}$  \\
$-3$ &$ \sqrt{\frac{2}{\operatorname{ln}4-1}-2}$  & $-\sqrt{\frac{2}{\operatorname{ln}4-1}-3}$ \\
$-4$ & $\sqrt{\frac{6}{5-3\operatorname{ln}4}-3}$ & $-\sqrt{\frac{6}{5-3\operatorname{ln}4}-4}$\\
$-5$ & $\sqrt{\frac{12}{6\operatorname{ln}4-7}-4}$ & $-\sqrt{\frac{12}{6\operatorname{ln}4-7}-5}$\\
\end{tabular}
\label{tab:prefactors negative ladders}
\end{table}

We now briefly return to the complex energies of the bulk wave function for $n\leq-1$. Similarly as described in the main text, we can construct an ansatz for the Hamiltonian of the form $ \psi^\pm_{\tau_zn,\text{bulk}}(x,k_y) =\left[\alpha^\pm_{\tau_zn} \phi_{n-1}(x,k_y),\beta^\pm_{\tau_zn} \phi_{n}(x,k_y) \right]^\mathsf{T}$, where $n\leq-1$. 
To obtain the values of $\alpha^\pm_{\tau_zn}$ and $\beta^\pm_{\tau_zn}$, we plug in the ansatz to yield the matrix

\begin{equation}
	\label{negative_hermite_hamiltonian}
	h_{\tau_zn} = \begin{bmatrix}
		A_{\tau_zn} & \FermiVelocity\sqrt{2 e \BField }d_{n}\\
		\FermiVelocity\sqrt{2 e \BField }c_{n-1} &  B_{\tau_zn}
	\end{bmatrix},
\end{equation}
where the diagonal terms are real numbers, and we remind $c_{n-1}d_n=n$. We solve the eigenvalue problem of this matrix to find the eigenvectors and eigenvalues. Due to the non-trivial nature of $c_{n-1}$ and $d_n$ for $n<0$, this matrix is not Hermitian, and it is reminiscent of the non-Hermitian Su-Schrieffer-Heeger model. This model describes a 1D nearest-neighbor tight binding chain on a bipartite lattice, where the hopping parameter from site 1 to site 2 is different from the hopping parameter from site 2 to 1. This asymmetry introduces non-Hermiticity in the Hamiltonian, and it has been found that, just as in our system for very negative $n$, the eigenvalues are in general complex-valued \cite{lieu2018topological}. Therefore, we conclude that due to the introduction of the edge in our system, the prefactors $c_{n-1}$ and $d_n$ are no longer hermitian conjugates for $n\leq 0$, as opposed to when $n>0$. This causes the matrix $h_{\tau_zn}$ to be non-Hermitian, allowing the energies of the bulk solutions corresponding to these $n$ to be complex-valued.

\section{Wave functions at discrete \texorpdfstring{$k_y$}{ky}}
\label{sec:wave functions at discrete ky}
In Sec.\ \ref{subsec:basis for wave function} we claimed that the wave functions $ \psi_{\tau_zn,\text{D-NLS}}$,$ \psi_{\tau_zn,\text{C-NLS}}$ and $\psi_{\tau_zn,\text{D-bulk}}$ ($n>0$) are good approximations at discrete values of the energy.
In this section we explain how to obtain the $k_y$'s corresponding to these energies, and show that this procedure indeed leads to good approximate wave functions.

We start by noting that since the wave functions under consideration are derived from (possibly non-normalizable) bulk states, we can attribute to each wave function the Landau energy of Eq.\ \eqref{eq:Landau Energies} that corresponds to the appropriate bulk state.
If this energy is complex, as is the case for the D-CLSs, we only look at the real part.
We denote these wave functions as $\psi^\pm_{\tau_zn,i}$ and
their corresponding energy as $E^\pm_{\tau_zn,i}$, where $i$ is an index that denotes the wave function class (D-bulk, D-NLS or D-CLS).
Since, presumably, the wave functions are good solutions at this energy, the expectation value of Hamiltonian \eqref{eq:htz shifted} must be equal to the same energy, and thus

\begin{equation}
\label{eq:self consistency Landau}
    \langle \psi_{\tau_zn,i}(x,k_y)| h_{\tau_z}(k_y)|\psi_{\tau_zn,i}(x,k_y) \rangle = E_{\tau_zn,i}.
\end{equation}
In this equation, the only unknown is $k_y$, which we can numerically solve for. 
This yields the specific $k_{y,\tau_z,i}$ at which wave function $\psi_{\tau_zn,i}(x,k_y)$ is a good solution to the Schrödinger equation.
Note that this method does not rely in any sense on discretizing and numerically solving the Schrödinger equation for the edge state.
As such, instead of numerical solutions, the wave functions should be thought of as analytic approximations or variational solutions.

\begin{figure*}[ht]
    \centering
    \includegraphics[width=0.9\linewidth]{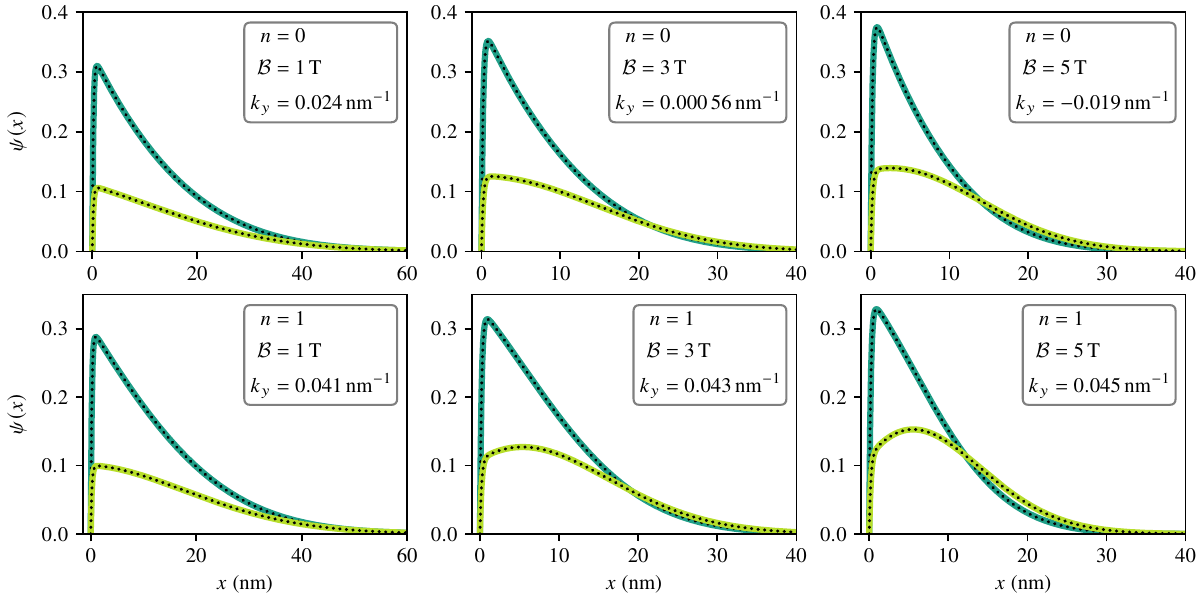}
    \caption{Wave functions of the edge state of \BiSe with $\tau_z = 1$ and several $\BField$ and $n$ at the corresponding Landau level energies as calculated by Eq.\ \eqref{eq:Landau Energies}. The panels with $n=0$ represent a D-NLS, and the panels with $n=1$ D-bulk states. The full lines are the numerically calculated wave functions, with the dark-green line being the first component and the light-green line the second. The dotted lines are the approximations of Eq.\ \eqref{eq:general_wave_function_edge}, with the $k_y$ corresponding to $E_n$ obtained via the self-consistency relation \eqref{eq:self consistency Landau}. It can be seen that the approximation works extremely well at all different values shown. We have found analogous results for $\tau_z=-1$.}
    \label{fig:A_Priori_Crossing_B135}
\end{figure*}

A comparison of the wave functions $\psi_{+1,1,\text{D-bulk}}$ and $\psi_{+1,0,\text{D-NLS}}$ for various magnetic fields is shown in Fig.\ \ref{fig:A_Priori_Crossing_B135}. 
It can be seen that the wave function is practically indistinguishable from the numerical solution.
We have found similar results for $\tau_z=-1$, and the wave functions $\psi_{\tau_zn,\text{D-NLS}}$.

It should be noted that for the wave function at certain energies, Eq.\ \eqref{eq:self consistency Landau} yields multiple solutions for $k_y$.
This can be understood by considering Fig.\ \ref{fig:Dispersion_edge_state_B1}. 
If we look for example at the positive Landau level with $n=2$, we see that, apart from the edge state, the energy is crossed by a higher energy band as well.
We have found that in this case the wave function $\psi_{\tau_z,2,\text{D-LS}}(x,k_y)$ does not only describe the edge state correctly, but also the higher lying state if we take the extra solution for $k_y$.
This was true for all $n$ that we checked (up to $n=10$) for both $\tau_z$.
An analytic function for all the energy bands should be obtainable via a procedure similar to that we applied for the edge state, but this lies outside the scope of this article.
For our purpose, we just need the insight that if multiple states cross a certain Landau energy, the crossing with the highest wave number corresponds to the edge state, as can be seen in Fig.\ \ref{fig:Dispersion_edge_state_B1}. 
Therefore, if Eq.\ \eqref{eq:self consistency Landau} returns more than one solution for $k_y$, it is this highest wave number that we identify as $k_{y,\tau_z n,i}$.

\section{Effect of lower-order D-CLS}
\label{sec:lower_order_D-CLS}

\begin{figure}[ht]
	\centering
	\includegraphics[width=0.8\linewidth]{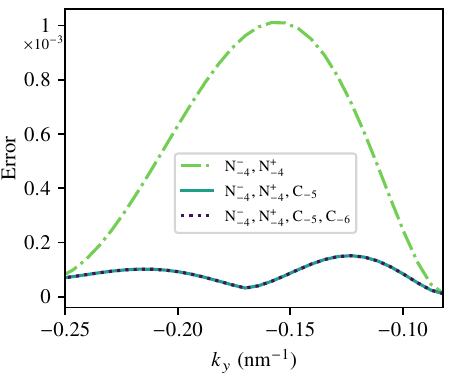}
	\caption{$L^{2}$ difference between the wave function $\psi_{\tau_z,\text{tot}} = \sum_{i \in I_1} c_i(k_y) \psi_{\tau_z,i}(x,k_y)$ and the numerical wave function for $\BField=\SI{1}{\tesla}$ and $\tau_z=1$, with different a number of wave functions included in the basis set \{$\psi_{\tau_z,i}$\}. Each set contains the functions $\psi^-_{\tau_z,-4,\text{D-NLS}}$ and $\psi^+_{\tau_z,-4,\text{D-NLS}}$, denoted by $\mathrm{N}^-_{-4}$ and $\mathrm{N}^+_{-4}$ respectively, and the different colors represent the different D-CLSs included, shown in the legend. The first set has no D-CLS, the second a single D-CLS of order $n=-5$, and the third two D-CLSs with orders $n=-5,-6$.  In contrast to, for example, the wave functions from Fig.\ \ref{fig:psi_error_B1}, the prefactors $c_i$, were not found by solving Eq.\ \eqref{eq:self consistency alpha}, but by directly fitting to the numerical wave function. It can be seen that adding a single D-CLS to the basis increases the accuracy of the wave function, but the effect of adding more is negligible.}
	\label{fig:comparison_D-CLS_multiple}
\end{figure}

In the main text we have introduced the Dirichlet complex Landau state (D-CLS) in Eq.\ \eqref{eq:wave function D-CL}. 
We note in Sec.\ \ref{subsec:Recap} that for the dispersion relation around $E=0$, the inclusion of just a single D-CLS yields an analytical approximation that performs excellently. 
We have found that if we include D-CLSs of lower order, the dispersion relation and its corresponding wave functions remain similarly accurate.
This is perhaps surprising, as one would expect that by adding additional basis vectors the wave functions can only become more accurate. 
This discrepancy is a result of the basis itself, and not an artifact of solving Eq.\ \eqref{eq:self consistency alpha}, as can be seen in Fig.\ \ref{fig:comparison_D-CLS_multiple}. 
This figure shows the error of the wave function $\psi_{\tau_z,\text{tot}} = \sum_{i \in I_1} c_i(k_y) \psi_{\tau_z,i}(x,k_y)$ where, in contrast to the main text, the parameters $c_i$ for a given $k_y$ are not found by solving Eq.\ \eqref{eq:self consistency alpha}, but are instead obtained by directly fitting to the numerical wave function.
It can be seen that in this case, the wave function performs much better after introducing a single D-CLS of order $n=-5$, and that the error tends to zero for $k_y = \SI{-0.17}{\nano\meter\tothe{-1}}$, which is the wave number that corresponds to the real part of the Landau energy of this D-CLS.
However, adding another D-CLS improves the accuracy only negligibly.
We have seen the same for wave functions with even lower D-CLSs, i.e. $n\leq-7$.
We suspect that this is the case because for increasingly low D-CLSs, the imaginary part of their Landau energy is increasing in magnitude. 
As a consequence, the assumption that these wave functions are good solutions at the wave number when the energy is equal to the (real part of) the Landau energy is likely to be incorrect. 
Hence, adding these D-CLSs to the total wave function does not increase the accuracy of the solution.

To summarize, we conclude that if one includes the highest order D-CLS to the solution, it increases the accuracy, but including lower order D-CLSs does not.
As including more basis wave functions makes the calculation of Eq.\ \eqref{eq:self consistency alpha} more complicated, we chose to only include a single D-CLS for all calculations in this article.

\newpage
\bibliography{ref}

@PREAMBLE{
 "\providecommand{\noopsort}[1]{}" 
 # "\providecommand{\singleletter}[1]{#1}%" 
}

@article{aguilera2019many,
  title={Many-body corrected tight-binding {H}amiltonians for an accurate quasiparticle description of topological insulators of the \ce{Bi2Se3} family},
  author={Aguilera, Irene and Friedrich, Christoph and Bl{\"u}gel, Stefan},
  journal={Physical Review B},
  volume={100},
  number={15},
  pages={155147},
  year={2019},
  publisher={APS},
  doi={10.1103/PhysRevB.100.155147}
}

@article{mong2011edge,
  title={Edge states and the bulk-boundary correspondence in {D}irac {H}amiltonians},
  author={Mong, Roger SK and Shivamoggi, Vasudha},
  journal={Physical Review B},
  volume={83},
  number={12},
  pages={125109},
  year={2011},
  publisher={APS},
  doi={10.1103/PhysRevB.83.125109}
}

@article{liu2010model,
  title={Model {H}amiltonian for topological insulators},
  author={Liu, Chao-Xing and Qi, Xiao-Liang and Zhang, HaiJun and Dai, Xi and Fang, Zhong and Zhang, Shou-Cheng},
  journal={Physical Review B},
  volume={82},
  number={4},
  pages={045122},
  year={2010},
  publisher={APS},
  doi={10.1103/PhysRevB.82.045122}
}

@article{zhang2014robustness,
  title={Robustness of quantum spin {H}all effect in an external magnetic field},
  author={Zhang, Song-Bo and Zhang, Yan-Yang and Shen, Shun-Qing},
  journal={Physical Review B},
  volume={90},
  number={11},
  pages={115305},
  year={2014},
  publisher={APS},
  doi={10.1103/PhysRevB.90.115305}
}

@article{zhang2009topological,
  title={Topological insulators in \ce{Bi2Se3}, \ce{Bi2Te3} and \ce{Sb2Te3} with a single {D}irac cone on the surface},
  author={Zhang, Haijun and Liu, Chao-Xing and Qi, Xiao-Liang and Dai, Xi and Fang, Zhong and Zhang, Shou-Cheng},
  journal={Nature physics},
  volume={5},
  number={6},
  pages={438--442},
  year={2009},
  publisher={Nature Publishing Group UK London},
  doi={10.1038/nphys1270}
}

@article{shan2010effective,
  title={Effective continuous model for surface states and thin films of three-dimensional topological insulators},
  author={Shan, Wen-Yu and Lu, Hai-Zhou and Shen, Shun-Qing},
  journal={New Journal of Physics},
  volume={12},
  number={4},
  pages={043048},
  year={2010},
  publisher={IOP Publishing},
  doi={10.1088/1367-2630/12/4/043048}
}

@misc{NIST:DLMF18.17.E4,
         key = "{\relax DLMF}",
       title = "{\it NIST Digital Library of Mathematical Functions}",
howpublished = {"\url{https://dlmf.nist.gov/18.17.E4}, Release 1.2.2 of 2024-09-15"},
         url = {"https://dlmf.nist.gov/18.17.E4"}
}

@article{rushka2020completely,
  title={A completely algebraic solution of the simple harmonic oscillator},
  author={Rushka, M and Freericks, JK},
  journal={American Journal of Physics},
  volume={88},
  number={11},
  pages={976--985},
  year={2020},
  publisher={AIP Publishing},
  doi={10.1119/10.0001702}
}

@article{lu2010massive,
  title={Massive Dirac fermions and spin physics in an ultrathin film of topological insulator},
  author={Lu, Hai-Zhou and Shan, Wen-Yu and Yao, Wang and Niu, Qian and Shen, Shun-Qing},
  journal={Physical Review B—Condensed Matter and Materials Physics},
  volume={81},
  number={11},
  pages={115407},
  year={2010},
  publisher={APS},
  doi={10.1103/PhysRevB.81.115407}
}

@article{zhou2008finite,
  title={Finite size effects on helical edge states in a quantum spin-{H}all system},
  author={Zhou, Bin and Lu, Hai-Zhou and Chu, Rui-Lin and Shen, Shun-Qing and Niu, Qian},
  journal={Physical Review Letters},
  volume={101},
  number={24},
  pages={246807},
  year={2008},
  publisher={APS},
  doi={10.1103/PhysRevLett.101.246807}
}

@article{kane2005z,
  title={$\mathbb{Z}_{2}$ topological order and the quantum spin {H}all effect},
  author={Kane, Charles L and Mele, Eugene J},
  journal={Physical Review Letters},
  volume={95},
  number={14},
  pages={146802},
  year={2005},
  publisher={APS},
  doi={10.1103/PhysRevLett.95.146802}
}

@article{zhang2010crossover,
  title={Crossover of the three-dimensional topological insulator \ce{Bi2Se3} to the two-dimensional limit},
  author={Zhang, Yi and He, Ke and Chang, Cui-Zu and Song, Can-Li and Wang, Li-Li and Chen, Xi and Jia, Jin-Feng and Fang, Zhong and Dai, Xi and Shan, Wen-Yu and others},
  journal={Nature Physics},
  volume={6},
  number={8},
  pages={584--588},
  year={2010},
  publisher={Nature Publishing Group UK London},
  doi={10.1038/nphys1689}
}

@article{neupane2014observation,
  title={Observation of quantum-tunnelling-modulated spin texture in ultrathin topological insulator \ce{Bi2Se3} films},
  author={Neupane, Madhab and Richardella, Anthony and S{\'a}nchez-Barriga, Jaime and Xu, SuYang and Alidoust, Nasser and Belopolski, Ilya and Liu, Chang and Bian, Guang and Zhang, Duming and Marchenko, Dmitry and others},
  journal={Nature communications},
  volume={5},
  number={1},
  pages={3841},
  year={2014},
  publisher={Nature Publishing Group UK London},
  doi={10.1038/ncomms4841}
}

@article{xia2009observation,
  title={Observation of a large-gap topological-insulator class with a single {D}irac cone on the surface},
  author={Xia, Yuqi and Qian, Dong and Hsieh, David and Wray, L and Pal, Arijeet and Lin, Hsin and Bansil, Arun and Grauer, DHYS and Hor, Yew San and Cava, Robert Joseph and others},
  journal={Nature physics},
  volume={5},
  number={6},
  pages={398--402},
  year={2009},
  publisher={Nature Publishing Group UK London},
  doi={10.1038/nphys1274}
}

@article{hsieh2009tunable,
  title={A tunable topological insulator in the spin helical {D}irac transport regime},
  author={Hsieh, David and Xia, Yuqi and Qian, Dong and Wray, L and Dil, JH and Meier, Fedorov and Osterwalder, J and Patthey, L and Checkelsky, JG and Ong, Nai Phuan and others},
  journal={Nature},
  volume={460},
  number={7259},
  pages={1101--1105},
  year={2009},
  publisher={Nature Publishing Group UK London},
  doi={10.1038/nature08234}
}

@article{qi2010quantum,
  title={The quantum spin {H}all effect and topological insulators},
  author={Qi, Xiao-Liang and Zhang, Shou-Cheng},
  journal={Physics Today},
  volume={63},
  number={1},
  pages={33--38},
  year={2010},
  publisher={AIP Publishing},
  doi={10.1063/1.3293411}
}

@article{linder2009anomalous,
  title={Anomalous finite size effects on surface states in the topological insulator \ce{Bi2Se3}},
  author={Linder, Jacob and Yokoyama, Takehito and Sudb{\o}, Asle},
  journal={Physical Review B},
  volume={80},
  number={20},
  pages={205401},
  year={2009},
  publisher={APS},
  doi={10.1103/PhysRevB.80.205401}
}

@article{maisel2024topology,
  title={Topology of \ce{Bi2Se3} nanosheets},
  author={Maisel Licer{\'a}n, Lucas and Koerhuis, S J H and Vanmaekelbergh, Daniel and Stoof, H T C},
  journal={Physical Review B},
  volume={109},
  number={19},
  pages={195407},
  year={2024},
  publisher={APS},
  doi={10.1103/PhysRevB.109.195407}
}

@article{liu2010oscillatory,
  title={Oscillatory crossover from two-dimensional to three-dimensional topological insulators},
  author={Liu, Chao-Xing and Zhang, HaiJun and Yan, Binghai and Qi, Xiao-Liang and Frauenheim, Thomas and Dai, Xi and Fang, Zhong and Zhang, Shou-Cheng},
  journal={Physical Review B},
  volume={81},
  number={4},
  pages={041307},
  year={2010},
  publisher={APS},
  doi={10.1103/PhysRevB.81.041307}
}

@article{vzutic2004spintronics,
  title={Spintronics: Fundamentals and applications},
  author={{\v{Z}}uti{\'c}, Igor and Fabian, Jaroslav and Sarma, S Das},
  journal={Reviews of Modern Physics},
  volume={76},
  number={2},
  pages={323},
  year={2004},
  publisher={APS},
  doi={10.1103/RevModPhys.76.323}
}

@article{hasan2011three,
  title={Three-dimensional topological insulators},
  author={Hasan, M Zahid and Moore, Joel E},
  journal={Annu. Rev. Condens. Matter Phys.},
  volume={2},
  number={1},
  pages={55--78},
  year={2011},
  publisher={Annual Reviews},
  doi={10.1146/annurev-conmatphys-062910-140432}
}

@book{bernevig2013topological,
  title={Topological insulators and topological superconductors},
  author={Bernevig, B Andrei},
  year={2013},
  publisher={Princeton University Press},
  doi = {10.1515/9781400846733}
}

@book{cage2012quantum,
  title={The quantum {H}all effect},
  author={Cage, Marvin E and Klitzing, Kv and Chang, AM and Duncan, F and Haldane, M and Laughlin, Robert B and Pruisken, AMM and Thouless, DJ},
  year={2012},
  publisher={Springer Science \& Business Media},
  doi={10.1007/978-1-4612-3350-3}
}

@article{mazumder2021brief,
  title={A brief review of \ce{Bi2Se3} based topological insulator: {F}rom fundamentals to applications},
  author={Mazumder, Kushal and Shirage, Parasharam M},
  journal={Journal of Alloys and Compounds},
  volume={888},
  pages={161492},
  year={2021},
  publisher={Elsevier},
  doi={10.1016/j.jallcom.2021.161492}
}

@article{pesin2012quantum,
  title={Quantum kinetic theory of current-induced torques in {R}ashba ferromagnets},
  author={Pesin, DA and MacDonald, AH},
  journal={Physical Review B},
  volume={86},
  number={1},
  pages={014416},
  year={2012},
  publisher={APS},
  doi={10.1103/PhysRevB.86.014416}
}

@article{moore2010birth,
  title={The birth of topological insulators},
  author={Moore, Joel E},
  journal={Nature},
  volume={464},
  number={7286},
  pages={194--198},
  year={2010},
  publisher={Nature Publishing Group UK London},
  doi={10.1038/nature08916}
}

@article{moes2024characterization,
  title={Characterization of the {E}dge {S}tates in {C}olloidal \ce{Bi2Se3} {P}latelets},
  author={Moes, Jesper R and Vliem, Jara F and de Melo, Pedro MMC and Wigmans, Thomas C and Botello-M{\'e}ndez, Andr{\'e}s R and Mendes, Rafael G and van Brenk, Ella F and Swart, Ingmar and Maisel Licer{\'a}n, Lucas and Stoof, Henk TC and others},
  journal={Nano Letters},
  volume={24},
  number={17},
  pages={5110--5116},
  year={2024},
  publisher={ACS Publications},
  doi={10.1021/acs.nanolett.3c04460}
}

@article{hasan2010colloquium,
  title={Colloquium: topological insulators},
  author={Hasan, M Zahid and Kane, Charles L},
  journal={Reviews of Modern Physics},
  volume={82},
  number={4},
  pages={3045--3067},
  year={2010},
  publisher={APS},
  doi={10.1103/RevModPhys.82.3045}
}

@article{yazyev2010spin,
  title={Spin {P}olarization and {T}ransport of {S}urface {S}tates in the {T}opological {I}nsulators \ce{Bi2Se3} and \ce{Bi2Te3} from {F}irst {P}rinciples},
  author={Yazyev, Oleg V and Moore, Joel E and Louie, Steven G},
  journal={Physical Review Letters},
  volume={105},
  number={26},
  pages={266806},
  year={2010},
  publisher={APS},
  doi={10.1103/PhysRevLett.105.266806}
}

@article{chiatti20162d,
  title={2{D} layered transport properties from topological insulator \ce{Bi2Se3} single crystals and micro flakes},
  author={Chiatti, Olivio and Riha, Christian and Lawrenz, Dominic and Busch, Marco and Dusari, Srujana and S{\'a}nchez-Barriga, Jaime and Mogilatenko, Anna and Yashina, Lada V and Valencia, Sergio and {\"U}nal, Akin A and others},
  journal={Scientific Reports},
  volume={6},
  number={1},
  pages={1--11},
  year={2016},
  publisher={Nature Publishing Group},
  doi={10.1038/srep27483}
}

@phdthesis{ThesisMoes2024,
  title        = {From atom to edge: {T}he influence of a magnetic field on nanoscale systems},
  author       = {Jesper R. Moes},
  year         = 2024,
  month        = {May},
  school       = {Utrecht University},
  type         = {Ph{D} thesis},
  doi = {10.33540/2237}
}

@article{shafiei2024tuning,
  title={Tuning the quantum phase transition of an ultrathin magnetic topological insulator},
  author={Shafiei, Mohammad and Fazileh, Farhad and Peeters, Fran{\c{c}}ois M and Milo{\v{s}}evi{\'c}, Milorad V},
  journal={Physical Review Materials},
  volume={8},
  number={7},
  pages={074201},
  year={2024},
  publisher={APS},
  doi = {10.1103/PhysRevMaterials.8.074201}
}

@article{shafiei2022controlling,
  title={Controlling the hybridization gap and transport in a thin-film topological insulator: Effect of strain, and electric and magnetic field},
  author={Shafiei, Mohammad and Fazileh, Farhad and Peeters, Fran{\c{c}}ois M and Milo{\v{s}}evi{\'c}, Milorad V},
  journal={Physical Review B},
  volume={106},
  number={3},
  pages={035119},
  year={2022},
  publisher={APS},
  doi = {10.1103/PhysRevB.106.035119}
}

@article{lieu2018topological,
  title={Topological phases in the non-Hermitian Su-Schrieffer-Heeger model},
  author={Lieu, Simon},
  journal={Physical Review B},
  volume={97},
  number={4},
  pages={045106},
  year={2018},
  publisher={APS},
  doi = {10.1103/PhysRevB.97.045106}
}
\end{document}